\documentclass[5p,twocolumn,number,authoryear]{elsarticle}
\usepackage{amsbsy}
\usepackage{amsmath}
\usepackage{amssymb}
\usepackage{verbatim}
\usepackage{fancyvrb}
\usepackage{xspace}
\usepackage{gensymb}
\usepackage{color}
\usepackage{enumitem}
\usepackage{listings}
\usepackage{placeins}
\usepackage{setspace}
\usepackage[draft=false]{hyperref}
\definecolor{DarkBlue}{rgb}{0.00,0.00,0.75}
\definecolor{LinkColour}{rgb}{0.063,0.55,0.85}
\hypersetup{
    linkcolor   = LinkColour,
    anchorcolor = LinkColour,
    citecolor   = LinkColour,
    filecolor   = LinkColour,
    urlcolor    = LinkColour,
    colorlinks  = true,
    pdftitle    = {A consistent approach to unstructured mesh generation for geophysical models},
    pdfauthor   = {Dr Adam S. Candy},
}
\usepackage{cleveref}
\providecommand{\norm}[1]{\left\|#1\right\|}
\providecommand{\eqref}[1]{(\ref{#1})}
\providecommand{\shingle}{Shingle\xspace}

\providecommand{\brepexpanded}{boundary representation\xspace}
\providecommand{\brep}{BR\xspace}\providecommand{\breps}{BRs\xspace}\providecommand{\brepexpandedacronym}{boundary representation (BR)\xspace}
\newcommand{\oned}{1D\xspace}\newcommand{\twod}{2D\xspace}\newcommand{\threed}{3D\xspace}

\providecommand{\zlevels}{$z-$levels\xspace}
\providecommand{\sigmalayers}{$\sigma-$layers\xspace}
\def\qad{\hskip0.6em\relax}

\newcommand{\podgpt}{\ensuremath{P_{1}^\mathrm{DG}-P_{2}}\xspace}

\crefname{equation}{}{}
\crefname{figure}{figure}{figures}

\newtheorem{constraint}{Constraints}

\crefname{constraint}{constraints}{constraints}

\crefalias{enumi}{tenet}
\crefname{tenet}{tenet}{tenets}

\newcommand{\constraints}{\cref{geobrep,geohmetric,geoidbound,geoidregion,geosurfbounds,geovmetric}\xspace}

\usepackage{titlesec}
\titlespacing\section{0pt}{5pt plus 4pt minus 2pt}{2pt plus 2pt minus 2pt}
\titlespacing\subsection{0pt}{5pt plus 4pt minus 2pt}{2pt plus 2pt minus 2pt}
\titlespacing\subsubsection{0pt}{5pt plus 4pt minus 2pt}{2pt plus 2pt minus 2pt}
\begin{document}
\setlength{\abovedisplayskip}{5pt plus 2pt minus 2pt}
\setlength{\belowdisplayskip}{5pt plus 2pt minus 2pt}
\begin{frontmatter}
\title{A consistent approach to unstructured mesh generation for geophysical models}
\author[tud,ese]{Adam~S.~Candy\corref{cor}}
\ead{a.s.candy@tudelft.nl}
\cortext[cor]{Corresponding author address:
Adam S. Candy, now at:
Environmental Fluid Mechanics Section, Faculty of Civil Engineering and Geosciences, Delft University of Technology, The Netherlands.%
}
\address[tud]{Environmental~Fluid~Mechanics~Section, Faculty~of~Civil~Engineering~and~Geosciences, Delft~University~of~Technology, The~Netherlands}
\address[ese]{Department of Earth Science and Engineering, Imperial College London, UK}

\begin{abstract}
Geophysical model domains typically contain irregular, complex fractal-like boundaries and physical processes that act over a wide range of scales.
Constructing geographically constrained boundary-conforming spatial discretizations of these domains
with flexible use of anisotropically, fully unstructured meshes is a challenge.
The problem contains a wide range of scales and a relatively large, heterogeneous constraint parameter space.
Approaches are commonly ad hoc, model or application specific and insufficiently described.
Development of new spatial domains is frequently time-consuming, hard to repeat, error prone and difficult to ensure consistent due to the significant human input required.
As a consequence, it is difficult to reproduce simulations, ensure a provenance in model data handling and initialization, and a challenge to conduct model intercomparisons rigorously.
Moreover, for flexible unstructured meshes, there is additionally a greater potential for inconsistencies in model initialization and forcing parameters.
This paper introduces a consistent approach to unstructured mesh generation for geophysical models,
that is
automated, quick-to-draft and repeat, and provides a rigorous and robust approach that is consistent to the source data throughout.
The approach is enabling further new research in complex multi-scale domains,
difficult or not possible to achieve with existing methods.
Examples being actively pursued in a range of geophysical modeling efforts are presented alongside the approach, together with the implementation library \shingle and a selection of its verification test cases.
\end{abstract}
\end{frontmatter}

\section{Introduction}
\label{sec:intro}

Numerical simulation models of geophysical processes employing flexible unstructured meshes
have advanced significantly.
In the field of ocean modeling
a mature class of models has evolved,
directed by regions of interest,
including:
ADCIRC \citep{westerink08}
for accurate basin-scale modeling of hurricane-induced storm surges,
FVCOM \citep{chen03} focused on coastal-scales,
H2Ocean \citep{cui13} applied to tsunami inundations,
SLIM \citep{hanertphd} and
D-Flow \citep{hagen14, dflowmanual} on rivers and marine estuaries,
QUODDY \citep{greenberg05} in the complex Canadian Arctic Archipelago,
T-UGOm \citep{lyard06} for improved tidal statistics,
Fluidity \citep{piggott08} for flexible non-hydrostatic studies
and
FESOM \citep{sidorenko11}
which has recently joined structured models in large, internationally coordinated climate model intercomparisons,
\citep[CMIP,][]{meehl07,taylor12}
and
\citep[CORE][\emph{Ocean Modelling} special issue]{coreii}.
These current approaches to
applying unstructured mesh methods
to
ocean modeling
are compared and considered in greater detail in \cite{danilov13}.

\begin{figure*}
\begin{center}
\includegraphics[width=\textwidth]{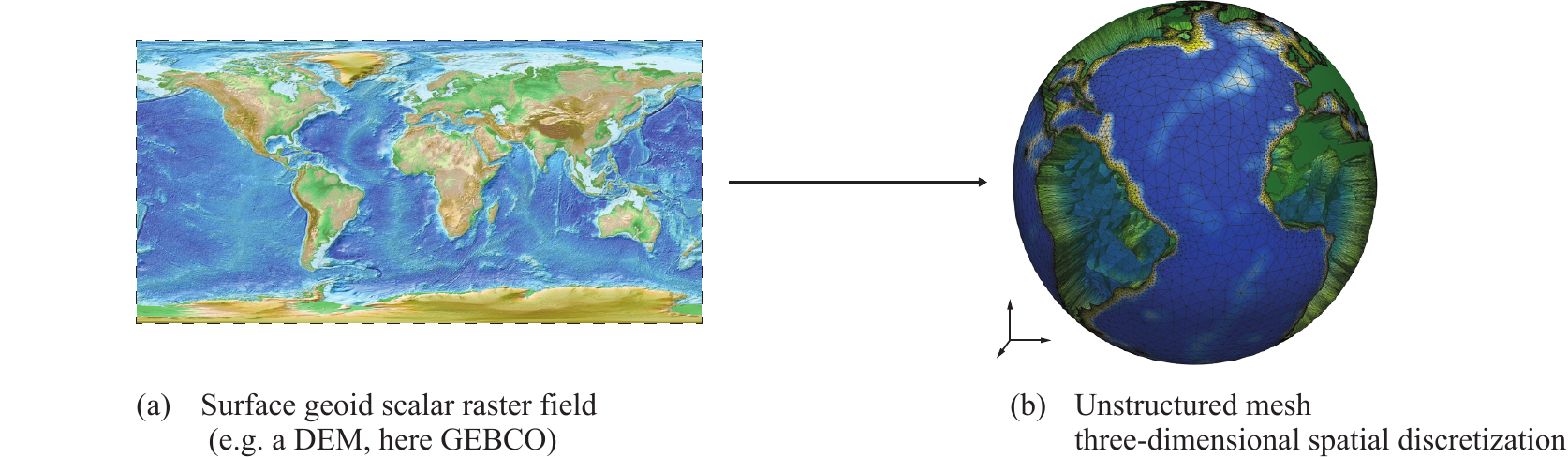}
\end{center}
\vspace{-3.8ex}
\caption{%
The challenge: to generate a domain discretization that is self-consistent,
through a deterministic, automated and repeatable process.
Geophysical model spatial discretizations are typically the product of multiple data sources and types (see \cref{fig:schematic}).
Fundamental to the approach here is the systematic development from a self-consistent data source,
such as
a surface geoid scalar raster field (a), e.g. the \cite{gebco} DEM,
to an unstructured mesh \threed spatial discretization (b).
}
\label{fig:challenge}
\vspace{-2ex}
\end{figure*}

Unstructured mesh approaches have the potential for distinct advantages over regular, structured grids.
Flexible conforming boundaries can accurately follow the complex, fractal-like bounds typically found in geophysical domains,
such as ocean coastlines and bottom bathymetry.
With local features exhibiting strong control over ocean dynamics \citep{danilov13c},
this can provide a more faithful, conforming representation.
Unstructured methods additionally support a flexible,
variable spatial resolution.
With ocean processes evolving over a diverse range of spatial and temporal scales
\citep[e.g. see][p55]{kantha00} --
from the large scale thermohaline circulation,
significant latitudinal variation in Rossby effects,
tides,
down to internal and gravity waves, double diffusion and oceanic turbulence
-- resolved spatial resolution
can be optimized
\citep[e.g.][]{sein16}
and efficiently capture a larger range of
spatial inhomogeneities to
avoid nesting,
using instead a seamless transition between large and small scales,
reducing also reliance on empirical parameterizations.
Moreover, unstructured approaches have the potential to be more computationally efficient as finer scales are included, with
\cite{holt17}
estimating a factor of 5--17 times less resources required.
Whilst unstructured mesh models may not replace structured modeling approaches completely, \cite{danilov13c} highlight there are certainly cases where this type of approach could be optimal.

The development of unstructured methods for ocean modeling have been actively discussed for the past fifteen years at the annual
\cite{imum}
workshops
\citep[][\emph{Ocean Modelling} special issue]{ham09},
with a focus on addressing problems in dynamical core discretizations.
Not only are the numerical discretizations of unstructured models more complex,
but
there exist
challenging hurdles
in their setup and initialization,
notably mesh generation.
Relative to regular gridded models,
which fundamentally require only a simple land mask to define domain bounds
with data remaining in a structured array format,
now a heterogeneous range of data types are required
and more advanced processing demanded.
This is an additional significant
barrier to wider adoption,
outside the specialist community.

\begin{figure*}[!h]
\vspace*{2mm}
\begin{center}
\includegraphics[width=\textwidth]{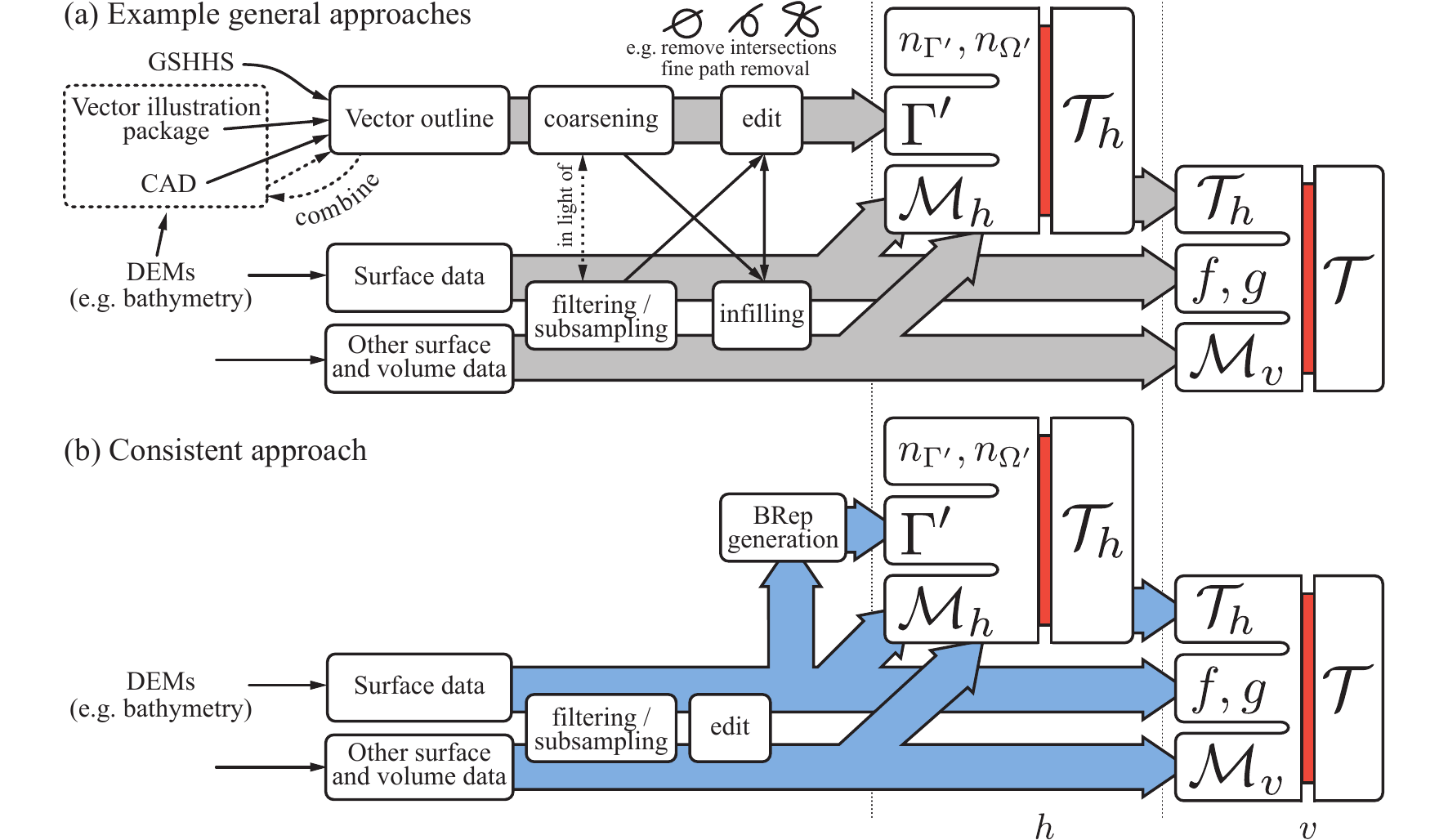}
\end{center}
\vspace{-3.8ex}
\caption{%
Approaches to unstructured mesh generation for geophysical models, highlighting
(a) the case in general
and
(b) the consistent approach introduced here,
with data pathways colored in grey and blue, respectively.
The connecting red sections represent spatial discretization processes $h$ \cref{h} and $v$ \cref{v},
that are fully constrained by the heterogeneous set of parameters provided to the left, and result in the output discretization to the right.
The former
$h$ provided by an established Delaunay triangulation implementation for example,
to generate the \twod tessellation $\mathcal{T}_h$.
With $v$,
for \threed discretisations,
an advancing front algorithm to extrude this tessellation $\mathcal{T}_h$ out to
$\mathcal{T}$.
}
\label{fig:schematic}
\vspace{-2ex}
\end{figure*}

Existing methods
\citep[reviewed in][]{candygis}
to construct underlying spatial discretizations
(which, for the purposes of the discussion here, specifically refers to the division of a continuous spatial domain into discrete parts -- a discrete tessellation or honeycomb -- a generalized notion of triangulation)
tend to be
model or application specific,
labor intensive and difficult to
reproduce,
or for a single purpose of testing to advance numerical discretization development.
These leave mesh discretizations difficult to edit and adjust,
when this flexibility is exactly one of the key advantages to unstructured mesh approaches.
Processes are often not well documented or accessible to those new to the field.
There is the demand for a
generic method, accessible to a wide range of modelers,
formalizing spatial discretization description for interaction with intercomparisons,
to support
and take full advantage of
the now growing class of mature unstructured mesh
numerical simulation models in the field.

The challenge is summarized in \cref{fig:challenge}:
to generate an unstructured spatial discretization
through a
deterministic and automated
process
from a set of self-consistent geoid surface geospatial data fields,
ensuring the self-consistency of data is propagated to the resulting spatial discretization (see \cref{fig:schematic}).
Domain bounds are becoming increasingly complex as simulations include a wider range of scales,
with \cite{holt17} predicting global models will include coastal scales down to 1.5km in the next 10 years.
With the range of scales, physics and required datasets diversifying,
it is a new and increasingly difficult challenge to ensure meshes and components in their construction are mutually consistent.

The objective of this paper is to provide:
\begin{enumerate}[leftmargin=1.2em]
\vspace{-1.35ex}
\setlength{\itemsep}{1pt}\setlength{\parskip}{0pt}\setlength{\parsep}{0pt}
\item A concise, formal description of the constraint problem (\cref{sec:constraint}, specifically \cref{constraint:geophysical}).
\item The solution requirements (\cref{sec:tenets}, specifically \cref{fig:tenets}).
\item Introduce a consistent approach to the generation of \brepexpanded to arbitrary geoid bounds (\cref{sec:brep,sec:metric,sec:id}).
\item Enabling rigorous unstructured mesh generation in general, for a wide range of geophysical applications,
in a process that is
automated,
quick-to-draft and repeat,
rigorous and robust,
and consistent to the source data throughout (\cref{fig:schematic}).
\end{enumerate}
This is implemented in the library developed as part of the project \cite{shingle},
and actively being applied in a range of current modeling studies, the details of some of which are  discussed in \cref{sec:application}.

The paper is structured such that the following
\cref{sec:meshinggeophysical} sets out a formal description of the problem in generating a discretization of geophysical domains and the challenge.
The new approach starts in \cref{sec:data} with self-consistent preparation of source datasets, followed by
accurate \brepexpandedacronym, spatial resolution and identification, in \cref{sec:brep,sec:metric,sec:id} respectively.
Generality of the approach is shown by the range of example applications, including verification cases,
presented in \cref{sec:application},
followed by conclusions.

\section{A complete description of domain discretization for computational geophysical models}
\label{sec:meshinggeophysical}
\subsection{Domain description}
Computational representations and manipulation tools of \threed objects
have traditionally been built up using a Constructive Solid Geometry (CSG) approach
in Computer Aided Design (CAD),
where Boolean operations are applied to primitive objects to develop the shape of the full domain, $\Omega\!\subset\!\mathbb{R}^3$.
These CAD based tools
have been extended to enable mesh generation within the bounds of
CSG defined objects
\citep[e.g.][]{comsol,gid,cubit}.
These approaches have been further extended to geophysical applications
with for example, GEOCUBIT \citep{geocubit} and
boolean operations on primitive objects to carve out an ice sheet in \cite{humbert09} using COMSOL.
Whilst these processes can be robust, and possible to automate to make less labor intensive,
it soon becomes a significant computational burden
for domain surfaces containing complex, multi-scale boundaries
which require an increasing number of \threed object intersection calculations.
Alternatively,
a domain
$\Omega$ is described in terms of its bounding surface $\Gamma$.
Just like the use of \emph{B-rep} descriptions
provide more flexibility in defining the curved surfaces of CSG objects,
this approach
offers the possibility to develop efficient descriptions that accurately follow
complex geophysical boundaries.

\begin{figure*}[!h]
\begin{center}
\includegraphics[width=0.95\textwidth]{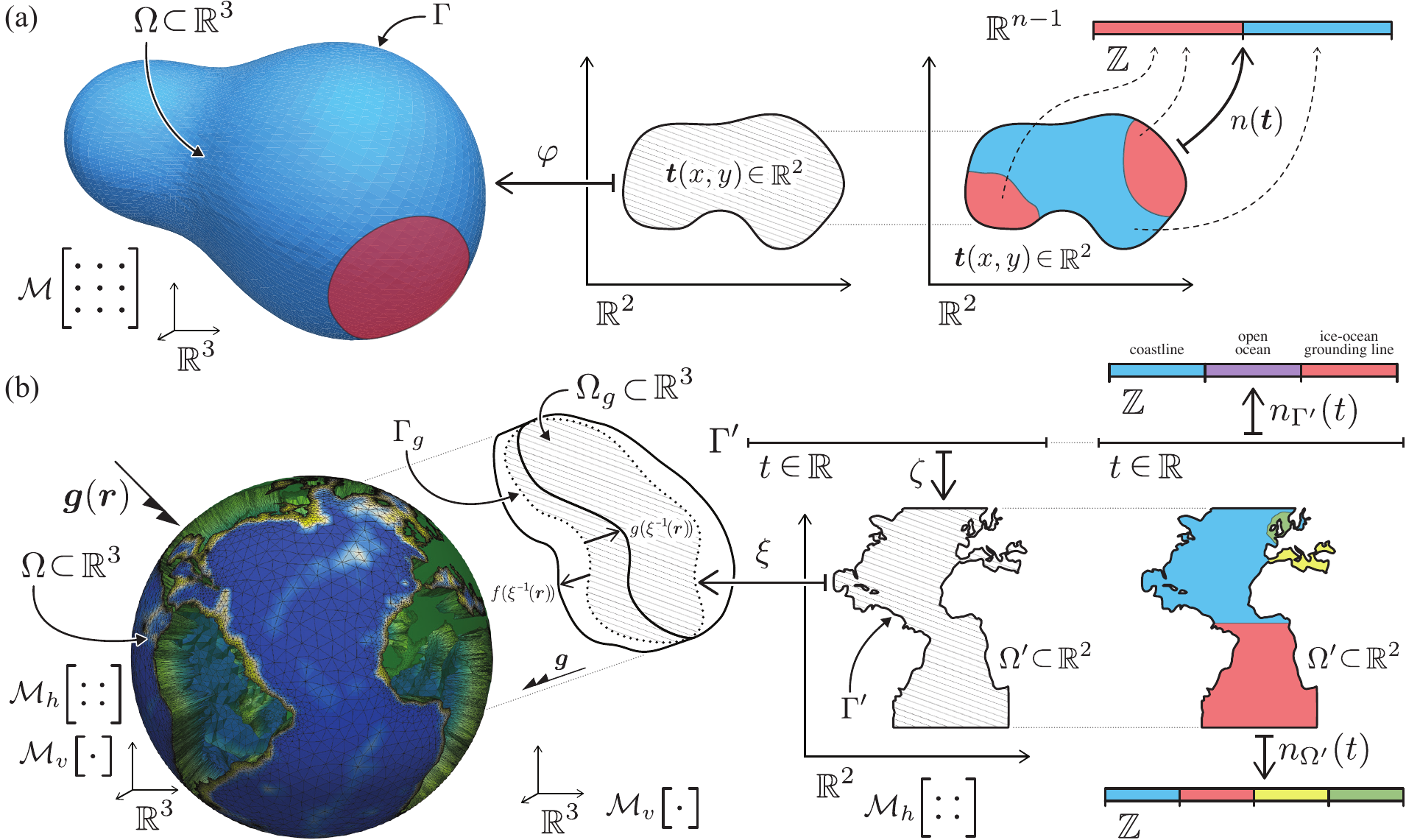}
\end{center}
\vspace{-2.8ex}
\caption{Breakout schematic of mesh generation for
(a) general unstructured spatial discretization
and
(b) typical geophysical domains,
referencing \cref{constraint:general,constraint:geophysical} of \cref{sec:constraint}.
}
\label{fig:breakout}
\vspace{-2ex}
\end{figure*}

\subsection{Constraints required to fully describe general domain discretizations}
The spatial domain discretization, or generation of meshes, for computational simulation in a domain $\Omega \subset \mathbb{R}^n$,
illustrated in the schematic (a) of \cref{fig:breakout},
requires constraining the following:
\begin{constraint}
\label{constraint:general}
\emph{(General spatial domain discretization):}
The spatial domain discretization for computational simulation in a domain $\Omega \subset \mathbb{R}^n$,
requires constraining a
\begin{itemize}[leftmargin=1.2em]
\renewcommand{\emph}[1]{\textbf{#1}}
\item \emph{Boundary representation} of the bounding surface $\Gamma \subset \mathbb{R}^{n-1}$,
an $n\!-\!1$ dimensional manifold,
defined using a parameterization
$\boldsymbol{t}$ under the homeomorphism
\begin{flalign}
\qad
\Gamma\!:
\boldsymbol{t} \in \mathbb{R}^{n-1}
\mapsto
\varphi(\boldsymbol{t}) \in \mathbb{R}^{n},
&&
\nonumber
\end{flalign}
including geometric constraints and the boundary identification
\begin{flalign}
\qad
n\!:
\boldsymbol{t} \in \mathbb{R}^{n-1}
\mapsto
n(\boldsymbol{t}) \in \mathbb{Z},
\textrm{ and an}
&&
\nonumber
\end{flalign}
\item \emph{Element edge-length resolution metric}, described by the functional
\begin{flalign}
\qad
\mathcal{M}\!:
\boldsymbol{x} \in
\Omega
\mapsto
\mathcal{M}(\boldsymbol{x}) \in
\mathbb{R}^n\!\times\!\mathbb{R}^n.
&&
\nonumber
\end{flalign}
\end{itemize}
\end{constraint}

\subsection{Spatial decoupling of the geophysical system}
\label{sec:decoupling}
Gravitational acceleration plays a dominant role in the evolution of geophysical systems,
with dynamics decoupled in locally orthogonal directions.
Buoyancy-driven effects force processes
in the local
\oned vertical direction aligned with gravity,
distinct from those
constrained to the \twod geoid plane ($\Omega_g$ in \cref{fig:breakout}).

This decoupling and significant difference in spatial and temporal evolution scales between the two motivate
numerical simulation models to treat these orthogonal directions differently.
Significant process has been made in unstructured ocean modeling based on depth-integrated equations
e.g. ADCIRC, FVCOM, H2Ocean, SLIM, T-UGOm) which calculate flow variation on the geoid plane.
Extensions to these calculate corrections to include non-hydrostatic effects in the orthogonal local direction
\citep[like the mode splitting employed by the MITgcm,][]{marshall97},
and in some cases solve in three-dimensions directly (e.g. Fluidity),
again with special consideration
of the
directional decoupling to
maintain hydrostatic \citep{ford04} and geostrophic \citep{maddison11} balances,
apply geometrically based multigrid \citep{kramer10}
and specifically treat the acute aspect ratios found in geophysical domains \citep{candy17}.

This decoupling motivates
a development of spatial discretization in parts,
in order to best support the associated dynamics,
considering first the geoid plane,
and secondly, if needed, an extrusion extending this in the normal direction.
The significant challenge in this problem is the spatial discretization of
the geoid plane, which is common to all unstructured numerical modeling approaches,
whether their dynamical cores require
two- or three-dimensional discretized domains.
As a result, this consideration and resulting approach
is applicable to all unstructured models of phenomena dominated by geophysical processes.

\subsection{Constraints required to fully describe geophysical model domain discretizations}
\label{sec:constraint}
In this case the constraints are increased,
with those of \cref{constraint:general} above being further divided,
such that mesh characteristics on the geoid plane are considered independently of those in the perpendicular direction of gravitational acceleration.
For domains in $\mathbb{R}^3$,
these become the following:
\begin{constraint}
\label{constraint:geophysical}
\emph{(Geophysical spatial domain discretization):}
The spatial domain discretization for a computational geophysics simulation in a domain $\Omega \subset \mathbb{R}^3$,
requires constraining a
\begin{itemize}[leftmargin=1.2em]
\renewcommand{\emph}[1]{\textbf{#1}}
\item \emph{Geoid \brepexpanded}
$\Gamma_g$,
of the geoid surface $\Omega_g \subset \mathbb{R}^3$ that results from the inverse prolongation operation
of the required domain
$\Omega \subset \mathbb{R}^3$
along the gravitational vector.
This is the maximal extent of the domain in the geoid plane.
Under a homeomorphism
$\xi$,
this is considered as
the chart
$\Omega' \subset \mathbb{R}^2$,
such that the boundary $\Gamma'$ is described by
\begin{flalign}
\qad
\Gamma'\!:
t \in \mathbb{R}
\mapsto
\zeta(t) \in \mathbb{R}^2,
&&
\label{geobrep}
\end{flalign}
an orientated vector path of the encompassing surface geoid bound defined in \twod parameter space, with a
\item \emph{Geoid resolution metric} for dynamics aligned locally to a geoid, described by the functional
\begin{flalign}
\qad
\mathcal{M}_h\!:
\boldsymbol{x} \in
\Omega'
\mapsto
\mathcal{M}_h (\boldsymbol{x}) \!\in
\mathbb{R}^2\!\times\!\mathbb{R}^2\!,
\textrm{ together with}\!\!\!\!
&&
\label{geohmetric}
\end{flalign}
\item \emph{Boundary and region identification}, prescribed by
\begin{flalign}
\qad
n_{\Gamma'}\!:
t \in \mathbb{R}
\mapsto
n_{\Gamma'} (t) \in \mathbb{Z},
\textrm{ and}
&&
\label{geoidbound}
\end{flalign}
\begin{flalign}
\qad
n_{\Omega'}\!:
\boldsymbol{x} \in \Omega'
\mapsto
n_{\Omega'} (\boldsymbol{x}) \in \mathbb{Z},
\textrm{ respectively,}
&&
\label{geoidregion}
\end{flalign}
\end{itemize}
gives the geoid `horizontal' domain discretization (a tessellation)
of $\Omega' \subset \mathbb{R}^2$,
with identification elements,
all denoted by
$
\mathcal{T}_h
$.
This together with
\begin{itemize}[leftmargin=1.2em]
\renewcommand{\emph}[1]{\textbf{#1}}
\item \emph{Surface bounds}, height maps defined on the surface geoid domain, described by the functions
\begin{flalign}
\qad
f,g\!:
\boldsymbol{x}
\mapsto
\mathbb{R}
\quad
\forall
\boldsymbol{x} \in \Omega',
\textrm{ and a}
&&
\label{geosurfbounds}
\end{flalign}
\item \emph{Vertical resolution metric} for dynamics in the direction of gravitational acceleration
(e.g. buoyancy driven),
described by the functional
\begin{flalign}
\qad
\mathcal{M}_v\!:
\boldsymbol{x} \in
\Omega
\mapsto
\mathcal{M}_v (\boldsymbol{x}) \in
\mathbb{R},
&&
\label{geovmetric}
\end{flalign}
\end{itemize}
gives the full domain discretization (of $\Omega \subset \mathbb{R}^3$) with identification elements,
all denoted by
$
\mathcal{T}
$.
\end{constraint}
This further restriction of constraints is illustrated in \cref{fig:breakout}(b),
and we note that
spatial discretization descriptions satisfying
\cref{constraint:geophysical} satisfy the more general \cref{constraint:general}.
See \cite{kramer10} for further details of the inverse prolongation operation referred to above.

In summary,
the horizontal discretization $\mathcal{T}_h$
is constrained by:
the
surface geoid domain $\Omega_g$ which is efficiently described in
\twod space as
$\Omega'$,
constrained by a \brep line $\Gamma'$ \eqref{geobrep}
parameterized under $t$;
the geoid element edge-length metric $\mathcal{M}_h$~\eqref{geohmetric};
together with
boundary and region identifications,
$n_{\Gamma'}$~\eqref{geoidbound}
and
$n_{\Omega'}$~\eqref{geoidregion}
respectively,
such that,
\begin{equation}
h\!: \{\Gamma', \mathcal{M}_h, n_{\Gamma'}, n_{\Omega'}\} \mapsto \mathcal{T}_h.
\label{h}
\end{equation}
The full discretization $\mathcal{T}$
of the \threed domain $\Omega\subset\mathbb{R}^3$,
is then constrained by:
this geoid discretization $\mathcal{T}_h$;
surface bounds $f(\boldsymbol{x})$ and $g(\boldsymbol{x})$~\eqref{geosurfbounds}
defined on $\boldsymbol{x}\in\Omega'\subset\mathbb{R}^2$,
that provide height extrusions
in directions parallel to
local $\boldsymbol{g}(\boldsymbol{r})$
for $\boldsymbol{r}\in\mathbb{R}^3$;
together with a
vertical edge-length metric $\mathcal{M}_v$~\eqref{geovmetric},
such that,
\begin{equation}
v\!: \{\mathcal{T}_h, f, g, \mathcal{M}_v\} \mapsto \mathcal{T}.
\label{v}
\end{equation}
A consequence of this development is that whilst spatial discretizations are unstructured in all three local coordinate directions, they are constrained to
ensure cell faces lie parallel or perpendicular to geopotential surfaces and dominant flow features,
such that fluxes can be calculated accurately
and
errors in the calculation of pressure minimized
\citep{danilov08,kramer10}.
This development is illustrated in the right hand side of the schematic in \cref{fig:schematic}, common to both (a) and (b).

\subsection{Challenges in domain discretization for geophysical models}
\label{sec:tenets}
Mesh generation for regular-gridded models is simply a matter of identifying which elements lie in the simulation domain through mask fields
and all data involved in the process is in the same \twod spatially-indexed scalar raster form,
using now standard operations for data on structured grid ocean models \citep{cotter08}.
For unstructured spatial discretizations,
the \cref{constraint:geophysical}:\constraints
require
a variety of data types,
from more standard \twod raster maps, to tensors and vector paths.
Notably, the geoid \brep \eqref{geobrep} is a vector path, in contrast to the \twod scalar data required for the majority of the rest of the constraints \cref{geohmetric,geoidbound,geoidregion,geosurfbounds,geovmetric}, and traditionally the raster forms used in the construction of structured grid meshes.
This makes it arguably the most challenging of the constraints to provide, ensuring it is an accurate and faithful representation that is consistent with \cref{geohmetric,geoidbound,geoidregion,geosurfbounds,geovmetric}.
Under this now heterogeneous set of constraints,
both the mesh description and generation problem are significantly more complex than the structured case
(summarized as the \emph{nine tenets of geophysical mesh generation} in \cref{fig:tenets}).

\begin{table*}
\centering
\begin{minipage}{\textwidth}
{
\noindent \hrulefill
\begin{enumerate}[
labelindent=*,
style=multiline,
leftmargin=*,
]
\vspace{-0.8ex}
\setlength{\parskip}{0pt}
\renewcommand{\emph}[1]{\textbf{\textsf{#1}}}
\item \label{tenet:brep}
      Accurate description and \emph{representation of arbitrary and complex boundaries}
      such that they are contour-following to a degree prescribed by the metric size field,
      with aligned faces so forcing data is consistently applied ($\Gamma'$, $f$, $g$).

\item \label{tenet:metric}
      \emph{Spatial mesh resolution} to minimize error; with efficient aggregation of contributing factors,
      ease of prototyping and experimentation of metric functions and contributing fields,
      complete over the entire extent of the bounded domain ($\mathcal{M}_h$, $\mathcal{M}_v$).
\item \label{tenet:region}
      Accurate geometric \emph{specification of regions} and \emph{boundary features};
      to provide for appropriate interfacing of regions of differing physics,
      model coupling and parameterization application ($n_{\Omega'}$, $n_{\Gamma'}$).
\item \label{tenet:consistent}
      \emph{Self-consistent}, such that all contributing source data undergoes the same pre-processing,
      ensuring self-consistency is inherited.

\item \label{tenet:efficient}
      \emph{Efficient drafting and prototyping} tools,

      such that user time can be focused on high-level development of the physics and
      initialization of the modeled system.

\item \label{tenet:scales}
      \emph{Scalability}, with operation on both small and large datasets,
      facilitating the easy manipulation and process integration, independent of data size.

\item \label{tenet:automated}
      \emph{Hierarchy of automation}, such that individual automated elements of the workflow
      can be brought down to a lower-level for finer-scale adjustments.
\item \label{tenet:provenance}
      \emph{Provenance} to ensure the full workflow from initialization to simulation and
      verification diagnostics are reproducible.
\item \label{tenet:standard}
      \emph{Standardization of interaction} to enable interoperability between both tools and scientists.
\end{enumerate}
\vspace{-1.8ex}
\noindent \hrulefill
}
\end{minipage}
\vspace{-1.8ex}
\caption{
The \emph{nine tenets of geophysical mesh generation},
that solutions to the spatial discretization of geophysical model domains should address.
}
\label{fig:tenets}
\end{table*}

Vector path descriptions are significantly more complex to store, interrogate and develop.
Control points of higher-order representations such as polynomial splines or flexible non-uniform rational B-splines (NURBS) do not necessarily lie on the bounding path.
Operations cropping to subregions, subsampling and merging are no longer simple selection, local binning routines, or efficient filtering matrix multiplications.
The path description should be an orientated vector generated to a required, spatially variable, level of accuracy (\cref{tenet:brep}) in a rigorous and reproducible (\cref{tenet:provenance}) manner.
In addition to ensuring paths accurately represent key geographic features (\cref{tenet:brep}) and
are efficiently stored
(optimized following algorithms such as \cite{douglas73}, for example)
it is important vector paths are topologically correct.
Line descriptions need to be closed to define fully bounded regions, that are disjoint and correctly orientated to identify which side of the path is to be included in the domain described.

\subsubsection*{Existing datasets of orientated vector paths}
\label{sec:existing}
To avoid issues in constructing these vector paths, modelers can use pre-prepared boundary datasets, such as the
Global Self-consistent, Hierarchical, High-resolution Geography Database,
\citep[GSHHG, previously GSHHS,][]{gshhs}.
Used as distributed, this data can be used successfully in model simulations.
The GSHHS plugin~\citep{legrand07,lambrechts08} written to interact with Gmsh~\citep{gmsh} successfully reads the database of pre-prepared coastline contours of GSHHS.
Significant progress has been made
in unstructured mesh ocean modeling with spatial discretizations generated using Gmsh
\citep[e.g. see ][]{legrand00,white08,scheltinga10,gourgue13,thomas14}.
This is a good solution for model problems with domains containing boundaries that can be defined by GSHHS,
although inconsistencies can develop when combined with other datasets,
to include additional geographic features, or
for the bounds \cref{geosurfbounds} for example.

Modifications, in practice, suffer from a lack of sufficient data, and additional external data is required to complete refinements.
It is also the case that
there are many domains for which a
vector boundary path is not available.
This has been mitigated to a small degree with the introduction of the updated GSHHG,
which additionally includes the CIA World Data Bank II rivers and border database,
but again in general this is very limiting.
This is the case in ocean domains extending under ice shelves to the grounding line where ice meets bedrock
or where it is important to extend over land to include the potential for inundation, in a tsunami model for example, or indeed in modeling geophysics of the past, in paleo-ocean simulations.
This is not a solution for arbitrary bounds (\cref{tenet:brep}) and demands an alternative approach.

\subsubsection*{Existing methods for \brepexpanded construction}
\label{sec:existingmethods}
Terreno \citep{gorman06,gorman07}
operates directly on DEMs, in line with \cref{fig:challenge}, combining \brep generation with optimization for shoreline and bathymetry representation.
This provides high-quality spatial discretization on a geoid,
but is limited in its flexibility and scope to add fine adjustments, and generally in its scope for a hierarchy of automation (\cref{tenet:automated}).

Vector illustration packages have long been used to handle
orientated vector paths and their editing.
These have matured over many years of development and contain robust interfaces
and efficient manipulation routines.
Interfacing with meshing software can be achieved through the standardized Scalable Vector Graphics (SVG) data type,
and this is the workflow applied together with Gmsh in
\cite{gourgue09},
\cite{brye11}
and
\cite{karna11},
and detailed further in \cite{gmshocean}.
New \breps for the domains of ancient seas were developed this way in \cite{wells10} following \cite{gorman08}.
Illustration tools however, have not been developed for this type of geographic processing,
and crucially do not consider projections of the sources or required output.
For instance, it is difficult to simplify path descriptions based on spatial distance in this approach.

A solution to this is demonstrated in \cite{candygis} using Geographic Information System (GIS) frameworks.
These are designed for \twod raster and polyline manipulations and importantly take into account dataset projections and geospatial information.
In highly multi-scale applications, this integration with these well-established mapping tools is a good, flexible and more rigorous solution for including intricate boundaries, over a range of scales, such as the man-made structures of a harbor together with the complex, fractal-like coastlines of the UK.
With this hand-editing and graphical approach, this is not the whole solution, since it can be difficult to automate (\cref{tenet:automated}) and reproduce (\cref{tenet:provenance}), and can become limited for complex, multi-scale boundaries,
but is an important part of a general geospatially informed solution approach.

\section{Self-consistent source data preparation}
\label{sec:data}
It is relatively easy to ensure a set of raster fields are mutually consistent (\cref{tenet:consistent}), with geospatially-aware matrix operations simply applied throughout.
Matched treatment of corresponding vector paths is a significant challenge.
Inconsistencies can develop,
with for example \breps lying over regions classified as land in the source bathymetry data, or worse bisecting other parts of the \brep.

\subsection{Automated generation of a consistent constraint description}
Central to the new, generalized approach of \cref{fig:schematic}(b) is that the whole discretized domain and forcing fields are built up from a self-consistent input dataset.
This dataset may contain multiple fields, but importantly they share a common spatial structure and have undergone the same harmonized processing (\cref{tenet:consistent}).
To ensure the process is efficient (\cref{tenet:efficient}), user interaction is focused on generating the consistently processed input dataset and high-level constraint description.
With the developed approach, it is then possible to automate subsequent processing
to lead to an output mesh and initialization described by the input,
an injective deterministic process.
This is repeatable and together with a record of the processing required for the input,
provides a complete record of provenance (\cref{tenet:provenance}).

\subsection{Data preparation and assimilation of datasets}
The preparation of data to describe the \cref{constraint:geophysical} alone contains many inherent challenges.
The \brep description \eqref{geobrep} should be a continuous, closed, non-intersecting path that is orientated and resolves well important selected geometric constraints (\cref{tenet:brep}).
The spatial resolution size descriptions \eqref{geohmetric} and \eqref{geovmetric}
need to be complete, defined over the entire geoid surface, appropriately graded so they vary smoothly enough so as not to adversely affect modeled dynamics
\citep{sein16,gorman06,piggott05}
and minimize numerical discretization error
(\cref{tenet:metric}).
Identification (\cref{tenet:region}) involves functions mapping over the range of spaces:
$\mathbb{Z}$,
$\mathbb{R}$ and
$\Omega'\subset\mathbb{R}^2$,
typically piecewise constant (e.g. finite element $\textrm{P}_0$) representations
on the discretized $\mathcal{T}_h$ and boundary $\Gamma'$
for $n_{\Omega'}$ and $n_{\Gamma'}$, respectively.
The heterogeneous set of constraint parameters and their discrete forms need to be kept mutually consistent in both the description preparation and mesh generation process (\cref{tenet:consistent}), whilst
achieving goals expected in scientific model development such as efficient prototyping, scalability, automation, provenance and interoperability (\cref{tenet:efficient,tenet:scales,tenet:automated,tenet:provenance,tenet:standard}).

\subsection{Resolution appropriate representation}
\label{sec:appropriate}
Source data is processed in order to ensure a good representation of fields and domain boundaries in the resulting discretization (see \cref{fig:schematic}).
A spatially inhomogeneous filtering focuses on areas of interest and provides support for physical phenomena.
For consistency, this must be applied equally to all sources, including vector paths, surface bounds and other surface and volume data.
This is not a trivial task, particularly in maintaining path consistency with geoid surface spatial datasets,
under this variable spatial resolution specification.

\subsection{Boundary representation preparation}
It is possible to encounter or, under the processing of \cref{sec:appropriate}, introduce intersections in path datasets
which require removal before \breps are passed on to meshing algorithms
(e.g. land boundaries passing through islands or loops in a single path, see \cref{fig:schematic}).
When hand-edited, a decision is made to separate the island,
adjust so they no longer intersect,
or simply remove the extra loop.
It is also possible that infilling of the spatial datasets is required~\citep{nurser} to ensure data is available in the region enclosed by these bounding paths.
This approach can be time-consuming, prone to human error and lead to inconsistencies.

\subsection{Iterative and incremental development}
In practice it is often found that a significant proportion of simulation failures for models on unstructured spatial discretizations are due to poor mesh quality \citep[e.g.][]{griffies00},
necessitating iterative, incremental changes to the underlying spatial discretization.
Additionally, unstructured meshes can contain errors in their construction that can be difficult to identify.
It is possible to introduce mesh elements that are free and decoupled from the rest of the domain, or due to the geometry and imposed boundary conditions, are fully prescribed from the outset, containing no independent parameters.
This stage often requires significant input from the user; to filter, subsample and hand edit,
together with other
preparatory stages shown in the context of the full mesh generation pipeline illustrated in \cref{fig:schematic},
and strongly motivates an approach which enables quick, efficient prototyping (\cref{tenet:efficient})
and a hierarchy of automation (\cref{tenet:automated})
for fine adjustments.

\section{Consistent generation of \brepexpanded constraint description}
\label{sec:brep}
\subsection{Boundary representation constraint data}
Constraint of the \brep requires an orientated vector path for \cref{geobrep}
and
for \threed models,
\twod scalar height maps complete within the surface geoid for \cref{geosurfbounds}.
Once the self-consistent source dataset has been prepared,
containing all data required to describe \cref{constraint:geophysical},
the rest of the process in \cref{fig:schematic}(b) is automated.
The first component is the generation of a suitable geoid \brep \eqref{geobrep} from this source dataset.

\subsection{Bottom-up, highest fidelity representation}
\label{bottomup}
At this stage, before requirements on the spatial resolution are considered, it is important the \brep is at the fidelity of the given source dataset.
This is a \emph{bottom-up} approach, in the same class as \cite{gorman07} and \cite{lambrechts08}, using the finest definition of the \brep polyline, such that it is then coarsened where possible.
This is in contrast to \emph{top-down} approaches to the approximation of domain bounds, such as \cite{douglas73},
which begin with a coarse definition and refine as required.
Whilst these can produce better results, they tend to be more expensive to compute.

Beginning with the highest-fidelity representation
means adjustment operations are kept local for computational efficiency
and can be relatively easily scaled in parallel.
It is easier to ensure the resultant discretized boundary is spatially consistent with the metric and vertical bounding fields,
starting with a path that is consistent.
Geophysical domain geoid surfaces are largely convex,
such that more flexible NURBS can be defined on the same control points as piecewise linear path representations
whilst maintaining consistency,
since the curves remain inside the convex hull of these points.
Lastly, with a \emph{bottom-up} approach,
the base description contains all information required to generate a full hierarchy of complexity in model domains,
such that it fully parameterizes $\Gamma'$, required for constraint \eqref{geobrep} and the functional $h$ \cref{h},
and thus can be shared for full reproducibility.

\subsection{Identifying the geoid \brepexpanded}
The self-consistent fields are
combined to form a mask identifying geoid bounds of the domain in directions parallel to geoid surfaces,
described by the functional
\begin{equation}
\boldsymbol{x} \in
\Omega'
\mapsto
\mathcal{F'} (\boldsymbol{x}, S_0(\boldsymbol{x}), S_1(\boldsymbol{x}), \ldots )
=
\mathcal{F} (\boldsymbol{x}) \in
\mathbb{R},
\label{pregeoidbrep}
\end{equation}
for source data
$
\boldsymbol{x} \in
\Omega'
\mapsto
S_i (\boldsymbol{x})
$, functions from $\mathbb{R}^2$ evaluating to variables of arbitrary rank and data type,
that are suitably
reduced by
the functional $\mathcal{F}$ to a scalar field.
It is the contour of this mask that
defines
the geoid \brep $\Gamma'$,
such that
\begin{equation}
\Gamma'\!:
\;\;
[t_0, t_1] \subset \mathbb{R}
\mapsto
\zeta(t) \in \mathbb{R}^2,
\
\textrm{where}
\;\;
\mathcal{F} (\zeta(t)) = c,
\label{geoidbrep}
\end{equation}
for a constant $c$.
In the case of a normalized mask centered about the boundary,
where
$\mathcal{F} \in [-1,1]$,
a constant value of $c = 0.0$ is taken.
For more common operations, a selection of forms for the functional $\mathcal{F}$ are available in the \shingle library, with
arbitrary functionals possible written directly as Python expressions.

The well-established and robust Generic Mapping Tools suite~\citep{gmtmanual} contains methods to generate contours from \twod raster fields, and could be used to solve \eqref{geoidbrep}.
It was found however, that the output GMT paths did not contain enough information to form a well-defined \breps on the geoid surface with distinct closed and open contours.
The process was also dependent on writing and reading multiple plain text files, which soon became inefficient as larger problems were considered.

We solve \eqref{geoidbrep} in the \twod parametric space of $\Omega'\in\mathbb{R}^2$,
under a homeomorphic projection $\xi$ that preserves neighbors
(see \cref{sec:chart}), in an approach built up from standard Python libraries.
In practice, in the scenarios presented in \cref{sec:application},
this is a cylindrical Mercator, stereographic or, over relatively small geoid patches, the Universal Transverse Mercator \citep[UTM, see][]{snyder87} projection.
Since there is no restriction this be isometric or area-preserving,
and moreover the preparatory stage may have intentionally yielded data at varying resolutions,
points in $\Omega'$
are not necessarily equally spaced or representative of target discretized resolution.
Spatial measures are calculated on $\Omega$, made under the transform to \threed Euclidean space and take into account curvature of the geoid surface.

Unlike existing approaches using GSHHS vector paths or attempting to use paths generated by GMT,
we have access to more information at this stage.
Alongside a description of polylines,
there is additionally access to
path orientation,
implied boundary IDs,
region IDs,
whether paths require closing,
physical boundaries,
and
imposed simulation boundaries, e.g. for simulation forcing.
The result is that the generation of the components required for $\mathcal{T}_h$ can be achieved consistently and largely automated, to minimize user edits and maintain a robust approach.

\subsection{Boundary closure}
\label{sec:closure}
The constraint~\eqref{geobrep} and associated resultant surface may:
(i) have a non-zero genus and contain island holes within the domain,
(ii) yield more than one closed contour path,
(iii) contain open convex and concave paths.
It is important these are handled automatically
to ensure \cref{tenet:efficient,tenet:scales,tenet:automated} are met.

For a simply connected surface with a zero genus and no island holes,
$\zeta(t_0) = \zeta(t_1)$
ensures the \brep is closed.
This is easily extended to non-simply connected surfaces with a non-zero genus with multiple intervals $[t_0, t_1] \subset \mathbb{R}$,
with a reversed vector path orientation denoting regions excluded from the surface.

At this stage open boundaries require closing to complete the domain such that it is consistent to metrics and surface bounds defined on the geoid, and identified correctly.
Boundaries are often closed along meridians and parallels, such that the domain is easy to specify and to facilitate model intercomparisons,
where, for example, forcings are provided on these closures.
This is relatively easy to achieve in structured models, where element faces typically lie on orthodromes, and often motivates their choice as bounds.
The approach depends on projections to different topological spaces for operations throughout the process
(through interaction with the established and robust \cite{proj4}).
Open boundary path closures are drawn under UTM, based on reference points local to the region, which are distance preserving and ensure minimal distortion.

\subsection{Path verification}
With the \emph{bottom-up} approach taken,
the finest resolution of the \brep is inherited from the source dataset, which itself has been prepared to be at the minimum goal fidelity, it is also possible to now check properties of the paths to eliminate features which could introduce problems at simulation time, e.g. an evaluation of path curvature, or a coarse check of the angle between successive segments.
Preliminary diagnostics on the \brep at this point provides direction for further processing of the source input dataset, and an iteration of this process to improve boundary selection and ultimately develop the best consistently generated mesh
\citep[pursued further in][]{candyshingle}.

With access to the highest resolution path data and associated projection information at this stage makes it possible to further automate additional processing, which is not possible or difficult with other approaches,
such as the explicit removable of islands by land area, or the automatic identification of inflows from river runoff.
In large multi-scale simulations there can be thousands of such features which makes processing time-consuming, error prone and severely impacts automation and the
efficient drafting and prototyping.
This is particularly important if the process illustrated in \cref{fig:schematic}(b) is to be repeated and iterated on.

\section{Consistent generation of spatial resolution constraint description}
\label{sec:metric}
\subsection{Spatial resolution constraint data}
\label{sec:spatial}
Spatial resolution is defined in two orthogonal components following the decoupling of \cref{sec:decoupling}.
This requires a functional \twod dyad, a rank 2 tensor field defined complete over the surface geoid $\Omega'$ for $\mathcal{M}_h$ \eqref{geohmetric}
and
for \threed models,
a scalar field defined complete over the whole domain $\Omega$ for $\mathcal{M}_v$ \eqref{geovmetric}.
The latter efficiently defined in a domain
bounded by
$\Omega'$, $f$ and $g$
(defined on $\xi^{-1}(\Omega)$, see \cref{fig:breakout}),
which determines the vertical coordinate system (see \cref{sec:hybrid}).

\subsection{Chart homeomorphism choice}
\label{sec:chart}
Location on the surface geoid $\Omega_g \subset \mathbb{R}^3$ is defined by two orthogonal linearly independent variables.
For efficiency of calculations and storage constraints
\cref{geobrep,geosurfbounds,geohmetric,geoidbound,geoidregion,geosurfbounds,geovmetric}
are defined on $\Omega' \subset \mathbb{R}^2$ under the homeomorphism $\xi$, which maps to
\threed space, and together form
the chart $(\xi,\Omega')$,
such that
\begin{equation}
\xi\!:
\boldsymbol{t} \in \Omega'
\mapsto
\xi(\boldsymbol{t}) \in \Omega_g \subset \mathbb{R}^{3},
\label{homeomorphism}
\end{equation}
where there exists a unique point $\boldsymbol{t}$ in $\Omega'$ for every point $\xi(\boldsymbol{t})$ on the surface geoid in the simulation domain $\Omega_g$, i.e. a bijective, invertible mapping so it is possible to move back and forth between chart and real \threed Euclidean space.
Additionally the mapping should be continuous, to preserve continuity of the surface geoid.

Where $\Omega'$ is also the domain required by the simulation
model (e.g. local UTM or longitude-latitude cylindrical
Mercator, such as ADCIRC)
post-processing of the output mesh is simplified,
although care is required to ensure $\xi$ in \eqref{homeomorphism} matches exactly that used within the model itself.
Where simulation calculations proceed in \threed Euclidean space $\xi$ is required as a post-processing step to map output to Cartesian coordinates in $\mathbb{R}^3$.
For simulation domains lying in a space distinct from $\Omega$ and $\Omega'$
(e.g. spherical polar coordinates) a further homeomorphic projection is required.
This does not affect the consistency of the approach, but permits a flexible choice of chart specific to  domain discretization and model simulation calculations.

In addition to working in a \twod chart, it is beneficial to choose a homeomorphism $\xi$ that is conformal
to minimize extremes in element anisotropy in the space
$\Omega'$
over which the \twod meshing algorithms operate.
For a global shell, a convenient conformal homeomorphism
is the stereographic projection with the
point antipodal to the center of projection removed
\citep[see][]{snyder87,lambrechts08,gorman06}.
The following form of stereographic projection is applied in combination with a standard spherical coordinate mapping to establish $\xi$
in the \cref{sec:application} applications to global oceans, ice shelf ocean cavities and Southern Ocean (SO)
\begin{align}
&
\xi\!:
(x, y) \in \mathbb{R}^2
\mapsto
(x, y, z) \in \mathbb{R}^3\!,
\ \textrm{with}\
\xi = \beta \alpha^{-1}\!,
\, \textrm{where}
\nonumber
\\
&
\qquad
\alpha(\psi, \phi, r)
=
2 r \, \mathrm{tan} \left( \frac{\pi}{4} - \frac{\phi}{2} \right)
(
  \mathrm{sin}\  \psi
,
 - \mathrm{cos}\  \psi
),
\nonumber
\\
&
\qquad
\beta(\psi, \phi, r)
=
r (
\mathrm{sin}\,\phi\ \mathrm{cos}\,\psi,\,
\mathrm{sin}\,\phi\ \mathrm{sin}\,\psi,\,
\mathrm{cos}\,\phi
).
\label{stereographicprojection}
\end{align}
To be bijective, the surface cannot include the projection point, which is propagated to infinity in the stereographic plane.
In the case of global ocean models this is achieved by slicing part of the land away at the South Pole to produce a truncated spherical shell that is homeomorphic to a single-point compactification of a \twod plane.
For smaller regional models on the sphere, this modification choice is more easily made.

Projections including multiple charts
or mappings that are not strictly homeomorphic can be handled with special treatment.
In the case of cylindrical Mercator,
the format typically used for global Earth datasets,
with $\Omega'$ cast in longitude-latitude space,
the topological atlas includes a homeomorphism that although conformal,
is not bijective and continuous along a meridian and at the poles.
\shingle automatically stitches together \brep vector paths broken across this edge meridian seam,
a fixing procedure made for GSHHS paths in \cite{lambrechts08}.
In the subsequent spatial discretization to this \brep description,
this issue is avoided using an azimuthal projection, such as stereographic \eqref{stereographicprojection}.

With the surface geoid \brep $\Gamma'$,
element metric $\mathcal{M}_h$,
and identifications $n_{\Gamma'}$ and $n_{\Omega'}$ generated following the consistent approach outlined above
to form the heterogeneous self-consistent constraint set
$\{\Gamma', \mathcal{M}_h, n_{\Gamma'}, n_{\Omega'}\}$,
corresponding to the
\cref{constraint:geophysical}:\cref{geobrep,geohmetric,geoidbound,geoidregion}
respectively,
the discretization $\mathcal{T}_h$ can be generated following the process $h$ (see \cref{h,fig:schematic}).
\shingle forms an accurate description of these first four constraints into a syntax
that can be interpreted by the meshing library
which in turn solves
the spatial discretization problem under these constraints.

\subsection{Interpretation and processing of the high fidelity constraint descriptions}
In the examples presented in \cref{sec:application},
\shingle prepares the constraint set in a syntax that
can be interpreted by the meshing library Gmsh~\citep{gmsh}
for discretization of the geoid domain $\Omega'$
in the process $h$ \cref{h}.
Gmsh is chosen because, through its standard syntax, it opens up access to a range of generic meshing algorithms that have been demonstrated robust with other approaches (\cref{sec:existing}).
It is equally possible to develop communication interfaces with other meshing packages and libraries, some of which can interact directly in Python (e.g. the \cite{trianglepython} to the Triangle library, \cite{triangle}),
discussed further in \cite{candyshingle}.
For a full \threed discretization, the remaining constraints \eqref{geosurfbounds}
and
\eqref{geovmetric}
are stored efficiently as fields on binary unstructured \cite{vtk} data structures containing a description of the spatial geoid discretization $\mathcal{T}_h$.
The first stage is a reparameterization of the
geoid \brep~\eqref{geoidbrep} produced by \shingle,
to take into account the mesh size field $\delta(\boldsymbol{x})$
derived from
$\mathcal{M}_h$, according to
\begin{equation}
\int_{t_0}^{\tilde{t}_i}
\frac{1}{\delta(\boldsymbol{x})}
\norm{
\frac{\partial \xi(t)}
{\partial t}
}
\,\mathrm{dt}
= i,
\quad  \textrm{for}\  i \in [ 0, 1, \ldots, n ],
\label{geoidbrepdis}
\end{equation}
for the $n+1$ new points along the boundary at parametric coordinates $\tilde{t}_i \in \{\tilde{t}_0, \ldots, \tilde{t}_n\}$,
with
$
\norm{
\partial \xi(t) /
\partial t
}
$
the length-scale
\scalebox{0.95}[1]{Jacobian of the homeomorphic parametric mapping $\xi$}.

The approach developed here and implemented within \shingle ensures
solution high-fidelity \breps calculated from \eqref{geoidbrep} are inherently non-intersecting.
The meshing algorithms within the Gmsh library do not natively maintain this
(with use of the adaptive trapeze rule for integrations to solve \eqref{geoidbrepdis})
but this has been addressed with the systematic recovery procedure described in \cite{lambrechts08}.
It is again important this is an automated algorithm, since it
cannot be handled rigorously and efficiently by hand when there is potential for multiple cases in the complex bounds of geophysical domains.
This provides an initial \oned computational mesh of the domain boundary $\Gamma'$ on the geoid.

The full computational meshing of the geoid surface $\Omega'$,
given the discretization of the boundary and geoid element metric $\mathcal{M}_h$,
is achieved by Gmsh through first an initial seeded Delaunay triangulation constrained to include the \oned boundary mesh.
This is then optimally restructured
using standard and robust approaches available in meshing libraries, here: an anisotropic Delaunay method \citep{george98}, a frontal algorithm \citep{rebay93} or a local modification technique \citep[see][]{gmsh,lambrechts08}.
The latter is similar to the method of mesh generation described in \cite{gorman06}, and the routines applied to adapt the mesh in time in response to solution dynamics, as described in~\cite{pain05}.
There is no need to reimplement these methods,
and instead we build on them and interact through standardized APIs and data structures.

\subsection{Boundary representation genus adjustment}
The geoid surface $\Omega'$ can be non-simply connected with a non-zero genus (\cref{sec:closure}).
For scales in the horizontal metric $\mathcal{M}_h$ larger than
these breaks in the surface,
the discretized boundary is coarsened from the high fidelity \brep following
\eqref{geoidbrepdis},
but
it is better to eliminate the \brep contribution entirely, and parametrize its influence.
In practice, meshing algorithms struggle to perform this elimination.
\shingle removes these features during the preparation of \eqref{geobrep}
by comparing their geoid extent to the metric $\mathcal{M}_h$ in its locality, or simply by a minimum area criteria.
Lengths and areas are calculated and compared in a projection that is local and appropriate to the intended simulation,
giving an accurate measure of distance on the geoid surface.

An alternative approach, also performed through \shingle in the examples of \cref{sec:application},
is to filter out these features through the filtering / subsampling stage shown in \cref{fig:schematic}(b),
and to a spatial resolution from
$\mathcal{M}_h$.
This leads to a more consistent set of constraints \{$\Gamma'$, $f$, $g$, $\mathcal{M}_h$, $\mathcal{M}_v$, $n_{\Gamma'}$, $n_{\Omega'}$\}
and is also how to handle groups of features, which can be agglomerated together at this
filtering / subsampling stage
to generate a larger-scale \brep and consistent \twod fields of the group together.

\subsection{Constraints over the number of degrees of freedom}
Spatial resolution is limited by the overall number of degrees of freedom and available computational resources, directly through available memory and indirectly by the cost of inter-node communications.
The maximum number of degrees of freedom can be a more natural constraint, rather than the element edge-length metrics of
$\mathcal{M}_h$
and
$\mathcal{M}_v$.
Unstructured mesh models permit a multi-scale heterogeneity in spatial scales within a single discretization and subsequent simulation.
This makes constraining on a maximum number of degrees of freedom no longer a simple arithmetic operation from a global spacing size.
Once the spatial pattern of element edge-lengths has been constructed as a functional of scalar fields, together with information of the order of representation made for prognostic simulation fields
that need to be stored in memory and potentially shared between nodes,
there is enough information to constrain on the maximum number of degrees of freedom.
The inverse of the functionals determining
$\mathcal{M}_h$
and
$\mathcal{M}_v$
are used to
calculate an approximate number of spatial nodes and degrees of freedom in
an output discretization,
which in turn is used as a constraint on these metrics.
This simply scales the element edge-lengths globally, or interacts in the functional to adjust the spatial pattern, limiting the smallest element edge-length for example, whilst the largest spatial spacing is maintained constant.

\subsection{Anisotropic spatial constraint}
The approach makes no restrictions over mesh element aspect ratios,
which are free to be fully anisotropic.
Anisotropy is strongly motivated in the orthogonal decoupled local horizontal and vertical directions by the differing physics characterizing geophysical systems highlighted in \cref{sec:decoupling}.
This is easily developed within the \cref{constraint:geophysical}
that describes and subsequently handles spatial discretization in these distinct directions in separate processes.
Within the geoid surface, anisotropy in the local plane orthogonal directions is prescribed by the tensor field
$M_h$ of \cref{geohmetric}.

\section{Identification of bounds and regions}
\label{sec:id}
Identification of regions on and within the surface boundary is required in order to apply geometric constraints and boundary conditions during a simulation,
providing \eqref{geoidbound} and \eqref{geoidregion} of \cref{constraint:geophysical}.
The former identifies edges of the domain $\Omega$ with normals orthogonal to local gravitational acceleration.
The remaining surfaces, that lie on the extruded geoid bounds defined by the scalar functions $f$ and $g$, are identified by the latter.
This second identification function also partitions the volume, for the application of simulation-time body forcings, viscosity parametrizations, vertical turbulence parametrizations and vertical coordinate systems, for example.

\begin{figure}[!h]
\begin{center}
\includegraphics[width=\columnwidth]{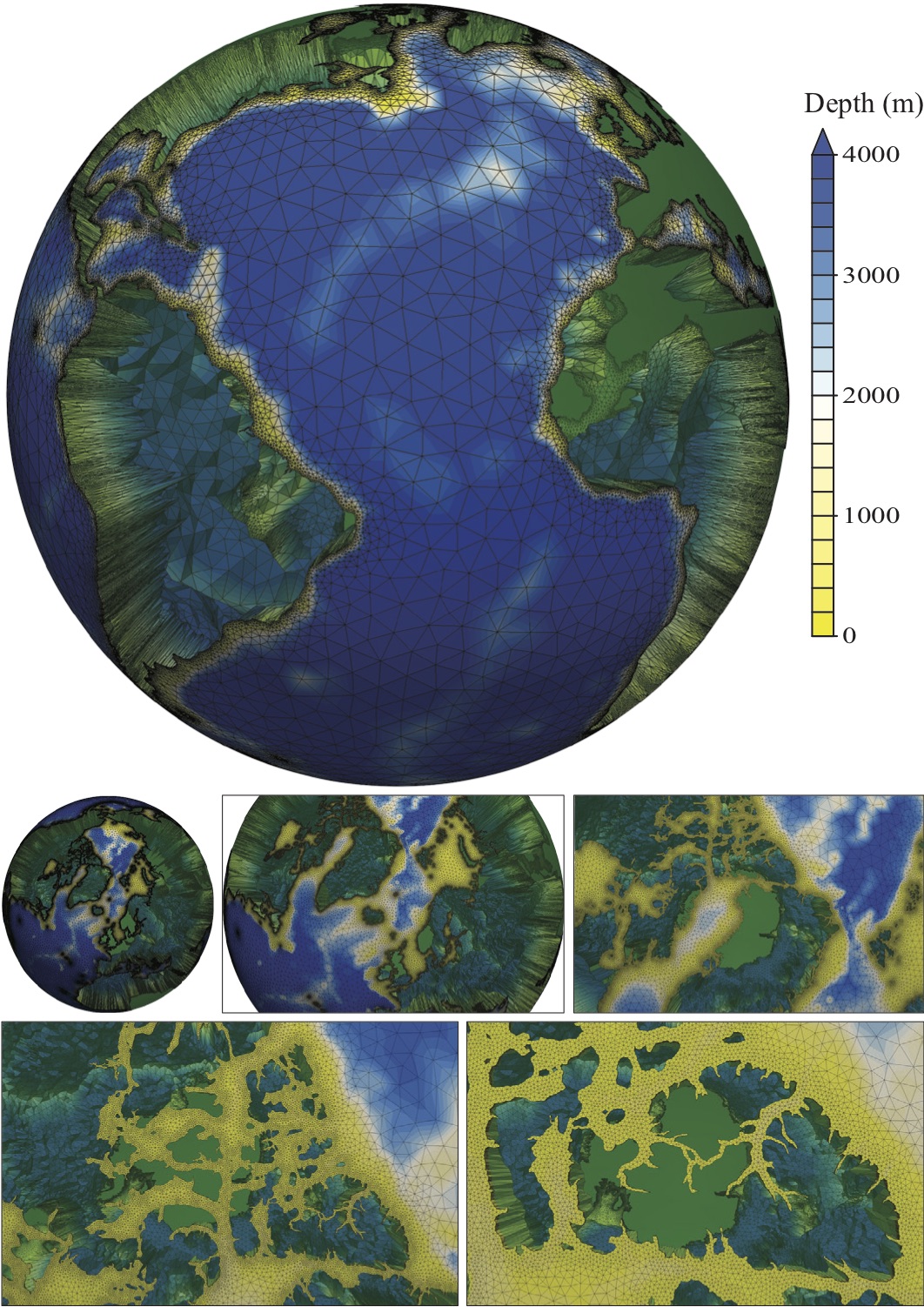}
\end{center}
\vspace{-3.8ex}
\caption{
Full mesh $\mathcal{T}$ of the global oceans containing a multi-scale of spatial resolutions, parallel to the geoid, from
$10\textrm{km}$
to
$500\textrm{km}$,
and vertical layers spaced from
$2\textrm{m}$
to
$500\textrm{m}$,
under differing regimes from
\sigmalayers in ice-covered and coastal regions up to the continental shelf,
transitioning to \zlevels in the open ocean.
The mesh contains 8,778,728 elements and 35,114,912 spatial degrees of freedom under its discontinuous Galerkin finite element discretization.
Zoomed in regions focusing on the complex
Canadian Arctic Archipelago
west of Greenland
around Ellesmere and Baffin island
are shown below.
The domain has been scaled radially by a factor of 300 in order to show the vertical extent of the discretization of this shell, with land regions shaded green.
}
\label{fig:global1}
\end{figure}

\subsection{Geometric constraints}
Geometric constraints apply physical restrictions on the domain discretization,
to ensure, for example, a land run-off outflow source to an ocean model is geographically placed correctly,
regions of differing drag are well-represented in an ice sheet or vegetation model,
or that the position of a terminating ice front is accurately maintained.
Geometric constraints
additionally optimize to the numerical discretization,
motivated by coupling differing models, nesting or matching to input data.
This identification is an integral part of the domain discretization
and is best developed and applied
while the high fidelity \brep is created, since it influences the placement of control points.

\subsection{Geoid surface closure}
Current approaches which require editing, constraint and identification of the \brep by hand are not appropriate for more complex boundaries
and an automated method is required.
Paths which fall outside the region of interest are truncated,
and closed along the edge of the region of interest
with joins geometrically constrained,
and by default identified differently to the main path internal to the domain so that appropriate boundary conditions can be applied.
Geometric constraint of the point of change joins ensure the discretized boundary is represented up to the edge of the region of interest, irrespective of the underlying local edge-element size provided by $\mathcal{M}_h$.

\subsection{Conforming extensions and closures}
For the purpose of model intercomparisons and benchmarks, the domain \brep is typically described in
terms of geographic contour sections, such as coastlines, bounded by parallels and meridians.
The approach automatically extends and closes domains along parallels and meridians, irrespective of projection.
The extension of the domain in this way can also be used to naturally include boundary restoring sponge regions in a geometrically consistent manner
to ensure these artificially created boundaries are accurate and well-represented, their interpolated positions are calculated in UTM space under a projection relative to a nearby point, and with an appropriate local step size,
with positions then mapped back to the required chart.

\subsection{Generalized hybrid vertical coordinates}
\label{sec:hybrid}
The identification also has direct input to the subsequent discretization stage that develops the extrusion to $f$ and $g$,
under $\mathcal{M}_v$,
and to match with nesting or coupled models.
This is used to develop a generalized hybrid vertical coordinate, smoothly varying between \zlevels and \sigmalayers \citep{haidvogel99,griffies05}
in the deep open ocean and coastal regions, respectively.
Additionally, a transition hybridized region is defined distinct in $n_{\Omega'}$, to limit reflection and rarefaction of waves whose propagation properties are dependent on spatial grid size.

\subsection{Demarcation of physical systems}
Vertical interfaces internal to the domain $\Omega$ and perpendicular to the geoid, are prescribed by a change in the region identification function
$n_{\Omega'}$.
Alternatively, if the interface is required to be geometrically constrained,
this is achieved through a definition of multiple
partially adjoining domains $\Omega_i$.
Both approaches are made in \cref{sec:application}.

The region identification $n_{\Omega'}$ additionally tracks the horizontal surfaces over which differing physical simulation models are coupled.
An ocean top surface interfaces with both air and ice, and $n_{\Omega'}$ is used such that
effects from ice--ocean interaction are only applied to a subset of the top ocean surface, e.g. melting and freezing processes, and loading from the ice above.
With information on the location of ice draft available in the source dataset \citep[as is the case in RTopo,][]{rtopo}, the identification can be made under the same treatments that are applied to the bounding surfaces (e.g. bathymetry and ice draft), to keep domain development self-consistent.

\section{Application case studies}
\label{sec:application}

A range of application examples are presented to emphasize the generality of the flexible and robust approach.
This begins with more standard domains such as the global oceans, that can be developed up to a point with other approaches using for example, the coastlines of GSHHG,
and proceeds to consider more complex domains bounded on all sides by geometrically intricate surfaces and containing multi-model coupled interfaces, such as ice shelf ocean cavity domains.
The basis constraint descriptions of these are available with the library, for use and further development.
\scalebox{0.97}[1]{Some additionally form part of the verification test suite.}

\begin{figure}[!h]
\begin{center}
\includegraphics[width=0.87\columnwidth]{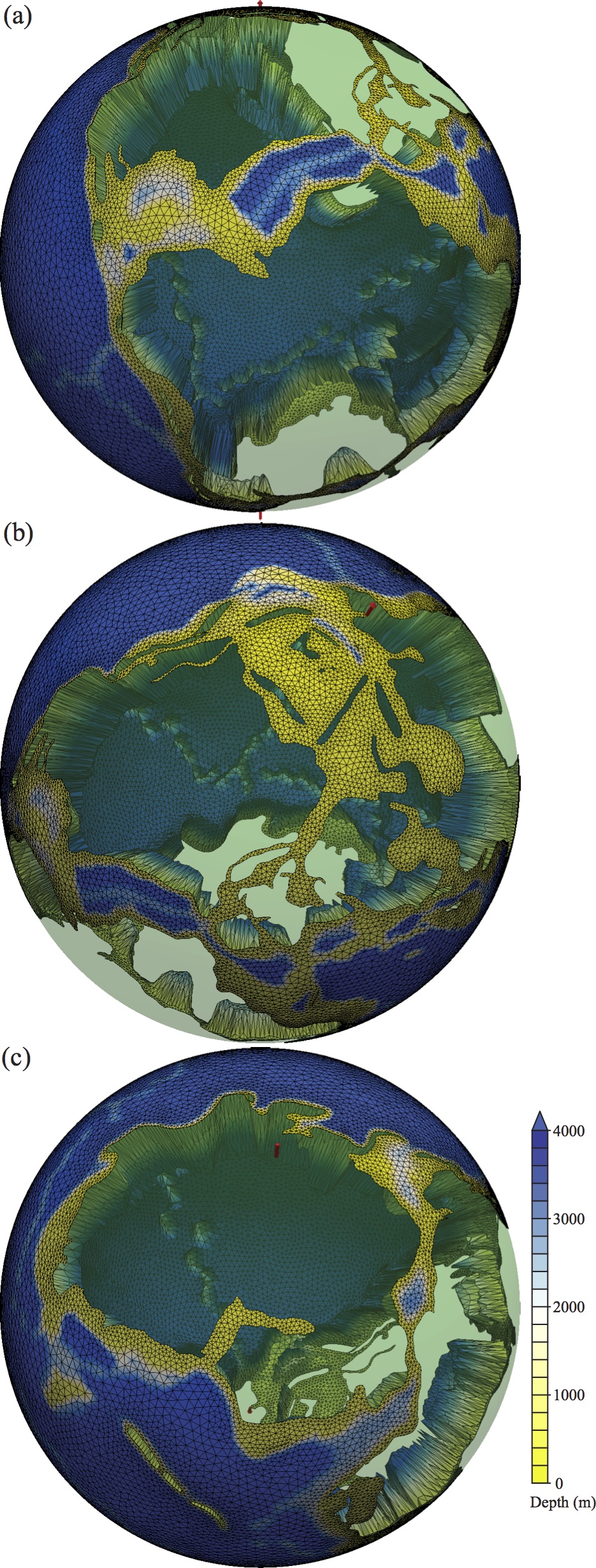}
\end{center}
\vspace{-3.8ex}
\caption{Full discretization of global oceans of the Early Cretaceous Berriasian age $\sim$140 million years ago, enabling research studies of paleo-oceans using variable resolution, boundary-conforming spatial discretizations,
that are particularly challenging during this period due to the
numerous shallow inland seas present from high eustatic sea levels.
The shell domain is radially stretched as per \cref{fig:global1}.
(a) Mid-Atlantic ridge central with North America above, West Gondwana below and the Pacific Ocean to the west.
(b) North America, Eurasia and the North Pole.
Tethys Ocean seen in the bottom right.
(c) East Gondwana and South Pole.
}
\label{fig:paleo1}
\vspace{-1ex}
\end{figure}

\subsection{Global oceans}
\label{sec:global}

Consistent spatial discretizations of the global oceans are developed from both the \cite{gebco} and RTopo \citep{rtopo} dataset sources, with the latter enabling the inclusion of water masses underneath the floating ice shelves.
With an approach and developed process that is
efficient, contains a hierarchy of automation and has a standardization of interaction
(\cref{tenet:efficient,tenet:automated,tenet:standard}),
it is a straightforward to switch source datasets.
From GEBCO, the discrete digital elevation map,
$\boldsymbol{x} \in \Omega' \mapsto d(\boldsymbol{x}) \in \mathbb{R}$,
is used as the
function $\mathcal{F}$
in \eqref{geoidbrep}, with $c = 0\textrm{m}$ marking the coastline.
In RTopo, there exists a field that identifies area type, which is consistent with the other fields provided, including importantly the depth.
It is therefore possible to generate the \brep \eqref{geobrep} from \eqref{geoidbrep} with $\mathcal{F}$ a function of this mask, used consistently with other constraints that are functions of different fields in the self-consistent dataset.

The geoid edge-length metric $\mathcal{M}_h$ \eqref{geohmetric} is a function of the depth field from the source dataset
$
d(\boldsymbol{x})\!: \boldsymbol{x} \mapsto d(\boldsymbol{x}) \in \mathbb{R}
$
and the proximity to coastline
$
p(\boldsymbol{x})\!: \boldsymbol{x} \mapsto d(\boldsymbol{x}) \in \mathbb{R}
$,
derived from the \brep
$\Gamma'$,
found using
the solution of a diffusion problem from the coastline boundary \citep[achieved easily through use of standard libraries such as the Geospatial Data Abstraction Library,][]{gdal}.
The isotropic geoid edge-length metric \eqref{geohmetric} demonstrated here is of the form
\begin{align*}
\mathcal{M}_{h}(\boldsymbol{x})
&=
\textrm{min}(
  \mathcal{M}_{g}(\boldsymbol{x}),
  \mathcal{M}_{p}(\boldsymbol{x})
)
\in \mathbb{R},
\;\;
\textrm{for}
\\
\quad
\mathcal{M}_g(\boldsymbol{x})
&=
\frac{
{10}^5
}{
3
}
\left({
\frac{
\mathrm{max}
(10, |d(\boldsymbol{x})|)
}{
10
}
}
\right)^{1/2}\!\!,
\\
\quad
\mathcal{M}_p(\boldsymbol{x})
&=
\left(
5\!\times\!{10}^5
-
{10}^4
\right)
\left(
\frac{
p - 3\!\times\!{10}^4
}{
2\!\times\!{10}^6 - 3\!\times\!{10}^4
}
\right)
+ {10}^4.
\end{align*}
This includes two factors guiding spatial resolution that are common in modeling ocean hydrodynamics.
The first
$\mathcal{M}_g$,
ensures gravity waves are accurately modeled and
the second
$\mathcal{M}_p$,
that coastlines are well-represented.
The form and number of components to the geoid metric are not critical to the demonstration, but that these are consistent with other constraints, including the \brep, and are efficiently combined in a robust and repeatable process
(\cref{tenet:consistent,tenet:efficient,tenet:standard}).

\begin{figure*}[!h]
\begin{center}
\includegraphics[width=0.9\textwidth]{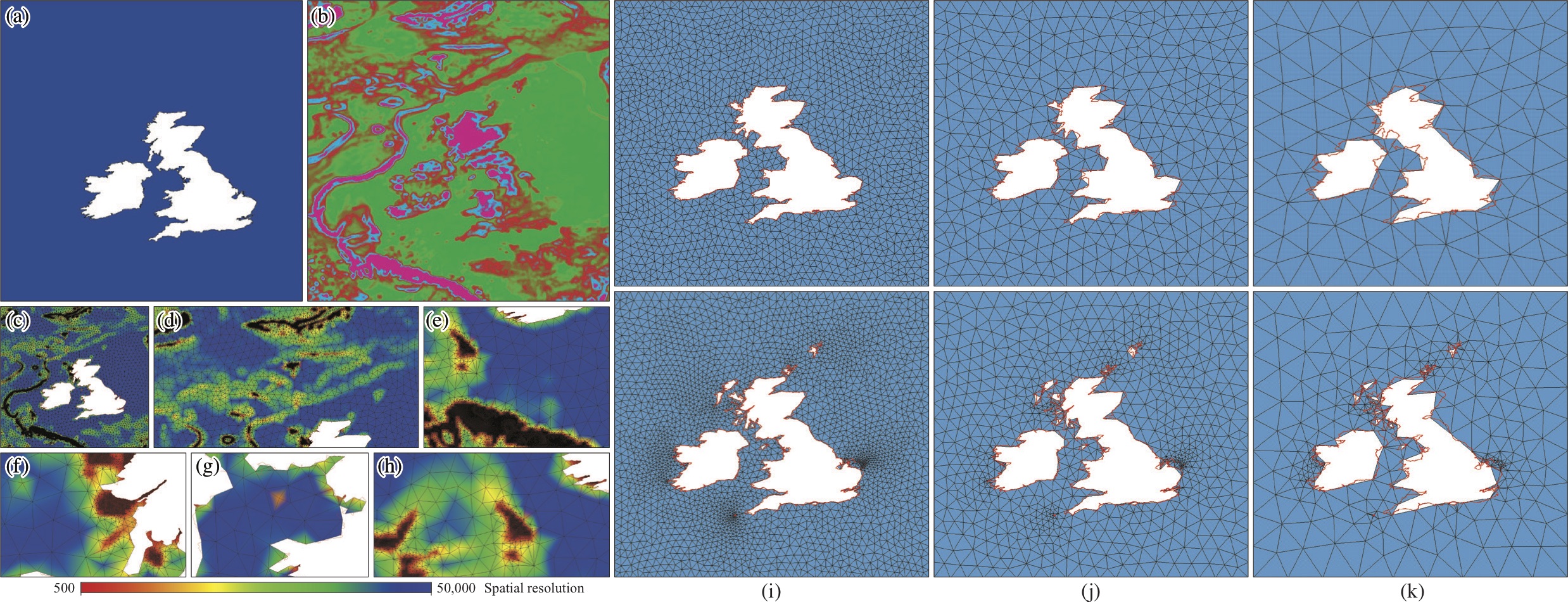}
\end{center}
\vspace{-3.8ex}
\caption{
Constraints on the discretization of a UK coastal seas domain containing the two largest land masses of the UK and Ireland, showing (a) a high fidelity \brep, and
(b) the associated metric $\mathcal{M}_h$ based on the sea bed gradient relative to the geoid plane.
The resulting spatial discretization, with local resolution highlighted in color, is shown for the full domain in
(c) and regions focused on:
(d) the Anton Dohrn seamount,
(e) the continental shelf and South West Approaches,
(f) Hebrides,
(g) Liverpool,
and
(h) Ireland.
At the successively coarser background resolutions of
(i) $10\mathrm{km}$,
(j) $25\mathrm{km}$ and
(k) $50\mathrm{km}$,
the mesh and discrete coastline generated through a \emph{bottom-up} approach from a high fidelity \brep described by \shingle and shown highlighted in red.
The top contains the two largest identified land features, and the bottom the first 49.
}
\label{fig:ukmetric}
\end{figure*}

To develop the full extruded domain
$\mathcal{T}$ (shown in \cref{fig:global1})
from
$\mathcal{T}_h$,
the surface bounds $f$ and $g$ are defined from depth and ice draft fields.
The vertical metric function
$\mathcal{M}_v$
implements a generalized hybrid vertical coordinate (see \cref{sec:hybrid}) and contains a specification of
$\sigma$-coordinates in ice-covered and coastal regions up to the continental shelf,
transitioning to \zlevels in the open ocean.
The extrude to achieve
$\{\mathcal{T}_h,\; f, g, \;\mathcal{M}_v\}
\mapsto \mathcal{T}$ is performed in parallel, with the
geoid discretization $\mathcal{T}_h$ divided by ParMetis and distributed in binary format across multiple processors and MPI processes.
The associated fields $f$, $g$ and $\mathcal{M}_v$ are similarly split, stored efficiently in binary unstructured \cite{vtk} data types and sent to the corresponding MPI processes.
This addresses scalability of \cref{tenet:scales}, facilitating mesh generation in parallel for geophysical domains and permitting the full discretization of very large multi-scale domains.

\subsection*{Global paleo-oceans}

In addition to developing a consistent discretization,
the motivation for this work is to enable automated generation to arbitrary geoid bounds.
The constraint description developed for \cref{fig:global1} is easily applied to the Rtopo dataset to include ice shelf ocean cavities in the hydrological domain.
Moreover, it is easily extended to
modeling ancient seas
in
domains with bounds that follow and conform to
coastlines.
Palaeoenvironment reconstructions of global bathymetry
utilizing geological observations
are used to build domains to ancient coastlines.
These are now simple, controllable and methodical modifications to the present day global ocean constraints (\cref{fig:global1}), and with the robust approach, easily applied for a range of ages
(\cref{fig:paleo1}).

\subsection{UK coastal seas}
\label{sec:uk}

A subset of GEBCO in the region
$
[-14.0,6.0] \times
[46.0,64.0]
$, for longitude-latitude coordinates ($\psi$,$\phi$)
is used to develop the spatial discretization of the UK coastal seas in \cref{fig:ukmetric}(a)--(h).
This raw source data
has a native resolution of 30 arc seconds, approximately $1\mathrm{km}$ on the geoid plane, and
this resolution is maintained for the boundary and metric constraint descriptions.
Following \cref{fig:schematic}(b) a high-resolution discrete approximation to \brep
\eqref{geobrep}
is generated by \shingle along the $0\mathrm{m}$ depth coastline
at this native resolution,
by solving \eqref{geoidbrep} with $d$ as the function $\mathcal{F}$ and the constant $c = 0$, i.e.
the path
$
\Gamma'\!:
[t_0, t_1] \subset \mathbb{R}
\mapsto
\xi(t) \in \mathbb{R}^2,
$
for
$
d(\xi(t)) = 0
$.
This \brep
$\Gamma'$
is shown outlined in \cref{fig:ukmetric}(a).
An isotropic geoid edge-length metric \eqref{geohmetric}
based on a measure of bathymetry gradient
of the form
\begin{equation}
\mathcal{M}_h (\boldsymbol{x})\!:
\boldsymbol{x} \in
\Omega'
\mapsto
\| \nabla d(\boldsymbol{x}) \|_2
=
\left(
\!
\left(
\frac{\partial d}{\partial x}
\right)^{\!\!2}
\!\!
+
\!
\left(
\frac{\partial d}{\partial y}
\right)^{\!\!2}
\right)^{\!1/2}
\!\!\!
\in
\mathbb{R},
\nonumber
\end{equation}
is developed in \cref{fig:ukmetric}(b) that is consistent with the \brep shown in \cref{fig:ukmetric}(a),
being also a function of the source GEBCO dataset $d$.

The same high fidelity discrete \brep
from \shingle
of
\cref{fig:ukmetric}(a) is paired with
a constant spatially homogeneous background geoid edge-length metric
\eqref{geohmetric}
of the form:
$
\mathcal{M}_h (\boldsymbol{x})\!:
\boldsymbol{x} \in
\Omega'
\mapsto
c
$, for a constant $c$,
to illustrate the \emph{bottom-up} approach
approximating the coastline at a consecutively coarser resolution in
\cref{fig:ukmetric}(i)--(k).
It is easy to edit contributions to the \brep with this approach, and \cref{fig:ukmetric}(i)--(k)
also presents meshes from a high fidelity \brep containing a larger number of islands.

\begin{figure*}[!h]
\begin{center}
\includegraphics[width=0.9\textwidth]{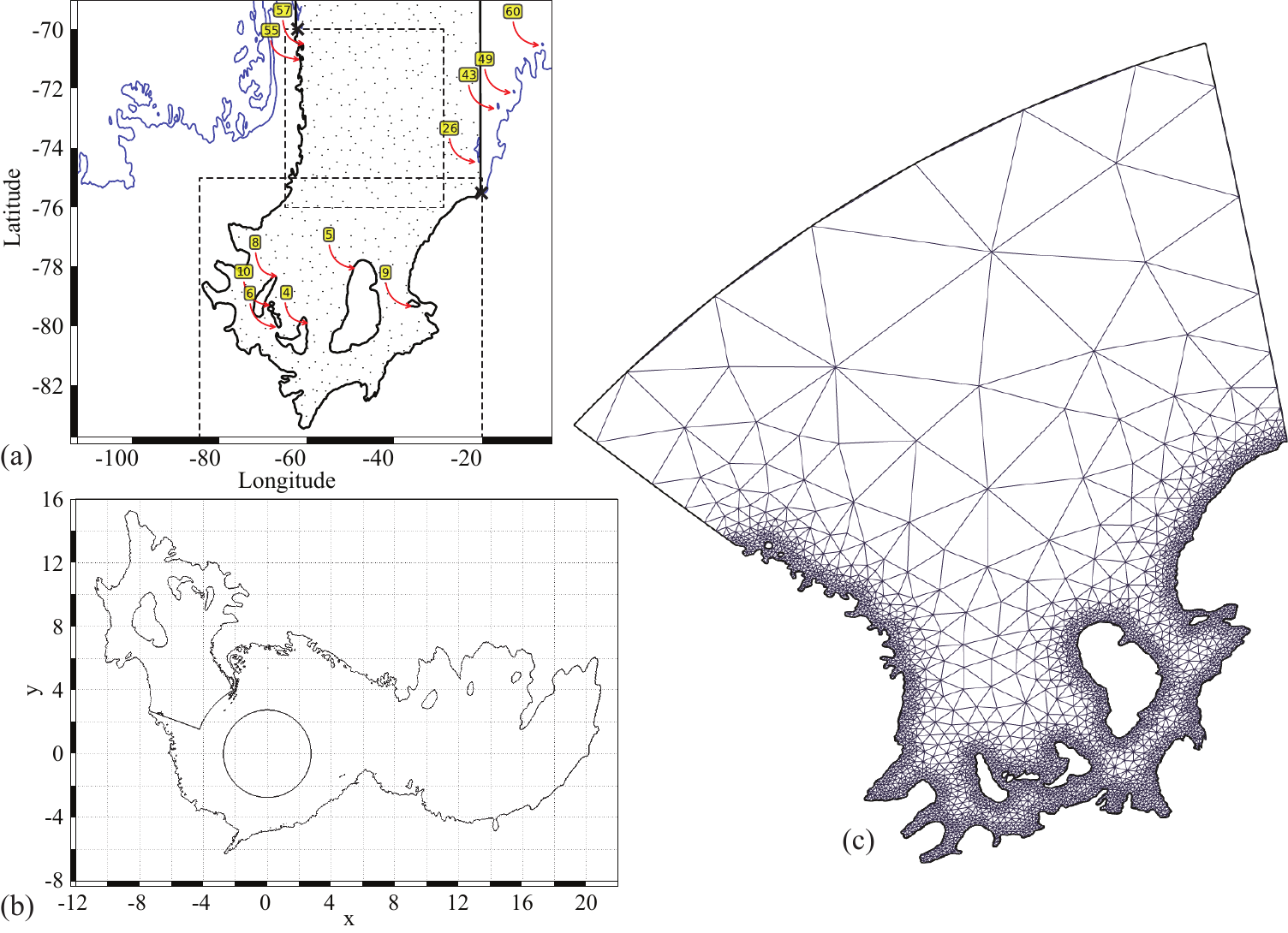}
\end{center}
\vspace{-3.8ex}
\caption{
(a)
Identification of the Filchner-Ronne ice shelf ocean cavity bounds from RTopo.
A total of 348 contours are identified in the SO upto a latitude of
50\degree S,
joined at the 180\degree W -- 180\degree E meridian,
and ordered by size.
For example, path 5 outlines Berkner Island that lies between the Filchner and Ronne ice shelf cavities.
The dashed lines mark the two specified regions of interest
and where paths exit their union
the boundary is extended by meridians up to
the specified
65\degree S latitude to be closed automatically by a parallel.
The figure shows direct graphical output, with
the two meridians and their points of intersection overlaid on top, together with the shading of the output domain, defined by the path orientation.
(b)
The \brep selected in (a) and generated by \shingle, shown under the stereographic projection~\eqref{stereographicprojection} about the North Pole,
in the chart under which the meshing algorithm operates,
together with the remaining grounding line contours identified in the SO and closure to the 60\degree S parallel.
(c)
Geoid surface mesh that results from the \brep development shown in (b), in \threed Euclidean space,
presented through an orthographic projection.
}
\label{fig:fr}
\end{figure*}

\subsection{Ice shelf ocean cavity}
\label{sec:iceshelf}

Domains including ocean cavities that sit below the floating ice shelves of Antarctica are bounded on the geoid plane
by grounding lines where ice meets bedrock, coastlines and the open ocean.
Immediately these present new challenges to automated, consistent discretization.
Grounding line positions are considerably harder to constrain than coastlines,
requiring observations from
Autonomous Underwater Vehicles (AUV), for example,
and tend to see frequent updates, with significant changes.
Orientated vector paths of these usually do not exist, or are soon out of date.
Additionally, unlike the global oceans, which are bounded only by coastlines, or the UK coastal sea example, which contained only an open ocean boundary,
this has a mix of boundary types and identifications.
Generation requires interaction with more bespoke datasets,
e.g. RTopo or finer resolution data direct from AUV observations.

\begin{figure*}[!h]
\begin{center}
\includegraphics[width=\textwidth]{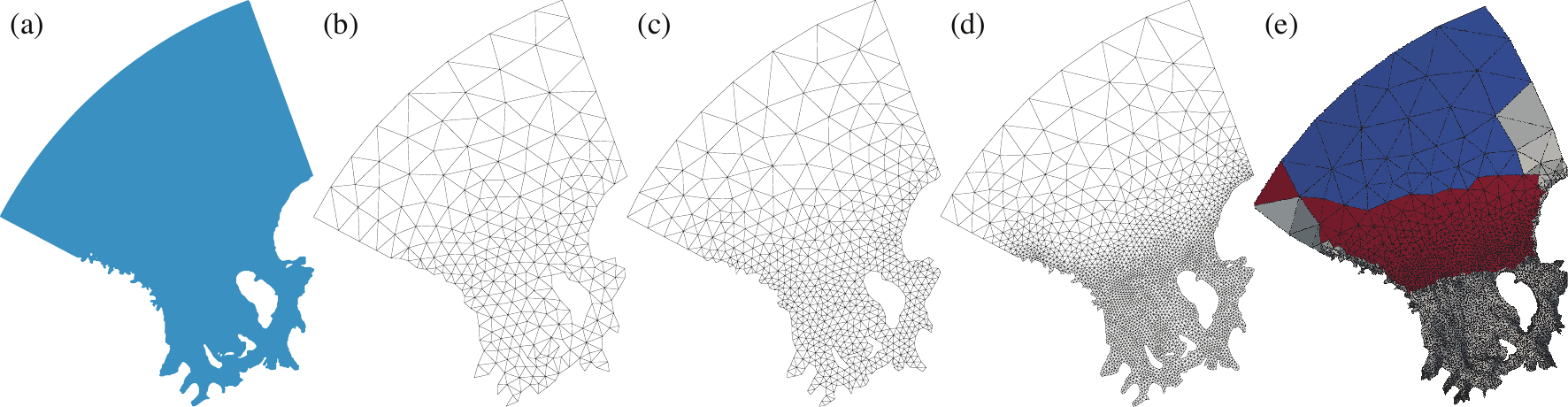}
\end{center}
\vspace{-3.8ex}
\caption{Filchner-Ronne ice shelf region containing the ten largest land masses.
(a) High-fidelity \brep $\Gamma'$ automatically constructed
from the grounding line up to the 65$^\circ$S parallel,
containing 18,354 control points at a spatial
resolution of 500m over a meridional extent of $\sim$1,600km.
This is automatically meshed, at three different background mesh metrics
$\mathcal{M}_h$ from 500km at the bounding parallel down to 100km, 50km and 20km at the grounding line
for (b) -- (d) respectively.
(e)
shows a
$\textrm{P}_0$ (element-wise)
identification function, for use in the extrusion to
$\mathcal{T}$.
All developed in \threed Euclidean space, presented through an orthographic projection about the South Pole.
}
\label{fig:fr1}
\end{figure*}

\subsubsection*{Filchner-Ronne ice shelf ocean cavity}
\label{sec:fr}

Filchner-Ronne is the second largest ice shelf in Antarctica (see \cref{fig:so2}),
approximately
$840 \textrm{km}$ long,
up to $600 \textrm{m}$ thick and
covering waters as deep as
$1,400 \textrm{m}$
at the grounding line.
Using the approach described, the domain containing the Filchner-Ronne ice shelf ocean cavity shown in \cref{fig:fr}(a) is straightforwardly captured with the
constraints presented in
\eqref{command}.
In this case the unfiltered self-consistent RTopo dataset for latitudes from 50\degree S to the South Pole is loaded
and limited to two selected regions described by bounding boxes
$[-85.0,-20.0] \times [-89.0,-75.0]$ and
$[-67.0,-30.0] \times [-76.0,-70.0]$, for longitude-latitude coordinates ($\psi$,$\phi$).

RTopo includes position data for the coast and grounding lines.
However, even those these are consistent with other spatial data within the source, this itself is not sufficient to derive a \brep, since points are not grouped, ordered, nor orientated to define the paths required to form \eqref{geobrep}.
Instead the derived region type mask \emph{amask} field of RTopo is used here directly to identify the ocean part of the surface geoid and a \brep that follows the grounding line below the floating ice sheet and coastline where cavities are not present.
As outlined in \cref{sec:data} above, the \emph{amask} field is filtered in the same way as other Rtopo fields, such as bedrock and ice draft, which are used later for vertical bound constraint.
The functional $\mathcal{F}$ of \eqref{pregeoidbrep} is formed efficiently with a modulo operation on the mask to combine regions identified as open ocean and ice-covered ocean (by integers 0 and 2, respectively), in contrast to the bare bedrock and grounded ice regions (by integers 1 and 3, respectively).

Where the paths are clipped by the bounding region, the domain is automatically extended with meridians up to the 65\degree S parallel and closed with boundaries identified as open ocean.
\Cref{fig:fr}(a) illustrates
one of the graphical interfaces
which can be used to display the paths identified and their unique label number with regions overlaid, for verification and more accurate selection when needed.
The resulting \brep is shown in \cref{fig:fr}(b),
and subsequent discretization in \cref{fig:fr}(c).

Through the \emph{bottom-up} approach,
\cref{fig:fr1}
highlights the ease at which it is possible to construct
finer and increasingly better resolved discretizations of the geoid boundary
generated by \shingle, in combination with Gmsh.
This makes it easy to both draft spatial discretizations themselves, using first coarse approximations during early prototyping stages,
and also in the development of a hierarchy of complexity in model simulations, where the level of detail captured is easy to control.

\begin{figure}[!h]
\begin{center}
\includegraphics[width=\columnwidth]{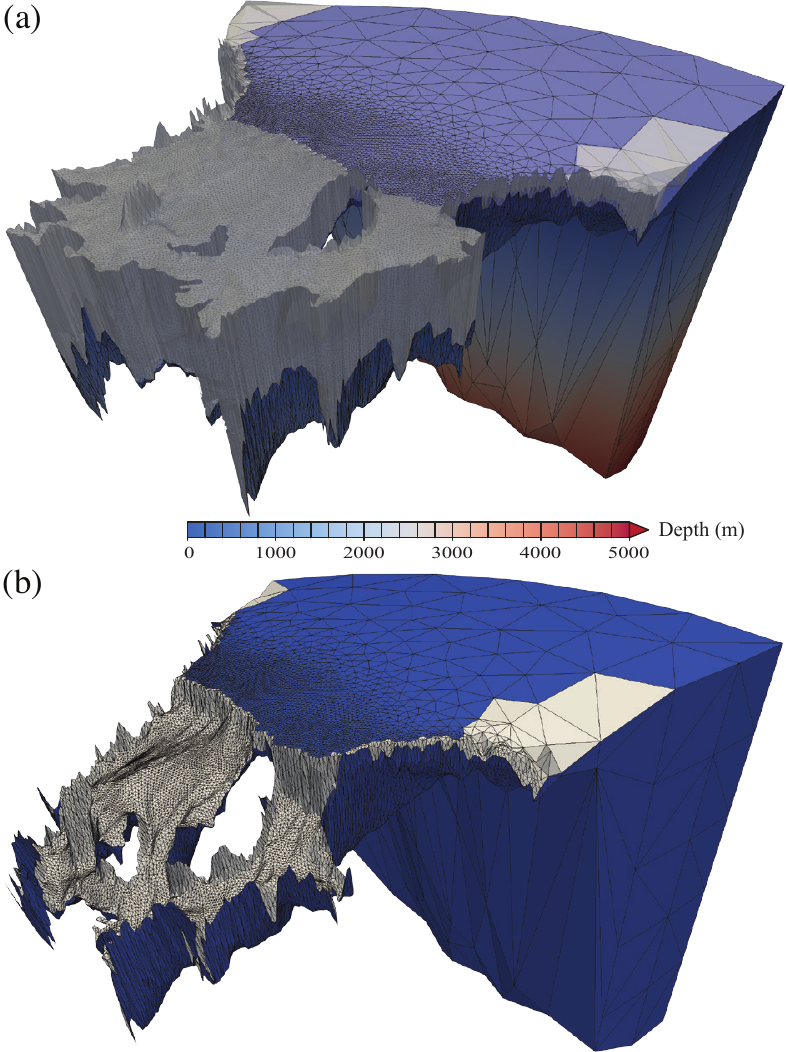}
\end{center}
\vspace{-3.8ex}
\caption{%
(a) Full discretized domains of the Filchner-Ronne ice shelf ocean cavity and floating ice sheet,
conforming to the complex external and internal geometric features and interface surfaces.
Vertical extent is radially exaggerated by a factor of 300.
(b) Identification function shown to pick out the melting and freezing coupling interface where the ocean meets the floating ice sheet.
}
\label{fig:fr2}
\end{figure}

In \cref{fig:fr2}(a),
the full discretized domains
of the ocean and floating ice sheets,
$\mathcal{T}^o$
and
$\mathcal{T}^i$
respectively
are shown,
with variable spatial resolution on the geoid.
The discrete domains
$\mathcal{T}^o$
and
$\mathcal{T}^i$,
with vertical bounds
$f^o, g^o$
and
$f^i, g^i$
respectively,
meet exactly at the ice--ocean interface,
with
\begin{equation*}
g^{o}(\boldsymbol{x})
=
f^{i}(\boldsymbol{x}), \quad \forall \; \boldsymbol{x} \in {\Gamma'}^{i}.
\end{equation*}

Incorporating these coupled interfaces in structured mesh models is relatively easy, and methods and implementations exists.
For unstructured-mesh models this is a significant challenge,
if one is to ensure the benefits of unstructured approaches are fully leveraged, with accurate conforming boundaries and multi-scale spatial resolution.

Geoid discretization is made on a plane through a stereographic projection to give a curved shell in \threed Euclidean space, which is then extruded to
the full discretization
$\mathcal{T}$.
The curvature of the representation of the
65$^\circ$S parallel in Euclidean space can be seen in
\cref{fig:fr2}.

\subsubsection*{Pine Island Glacier ice shelf ocean cavity}
\label{sec:pig}

\begin{figure}[!h]
\begin{center}
\includegraphics[width=0.9\columnwidth]{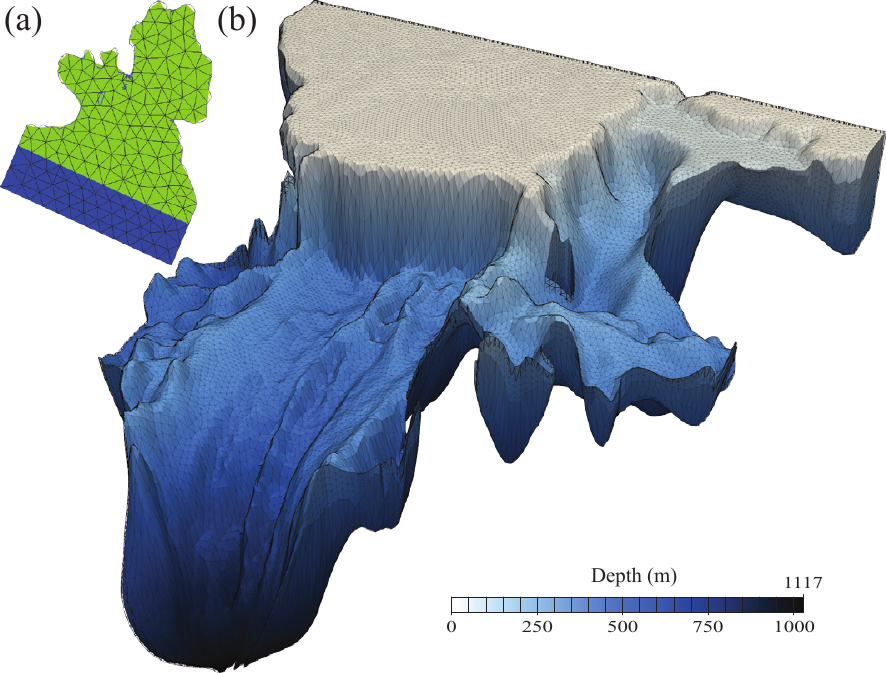}
\end{center}
\caption{
Pine Island Glacier ice shelf ocean cavity, with
(a) presenting the consistent high fidelity \brep constructed to conform to the grounding and coastlines marked by the self-consistent source dataset.  A relatively coarse spatial approximation is shown, with an additional sponge region added and shown in blue.
(b) Full discretized domain, at a finer homogeneous geoid resolution, inheriting the self-consistency of the source data fields.
}
\label{fig:pig1}
\end{figure}

Pine Island Glacier (PIG) ice shelf ocean cavity
is located in the Amundsen Sea region of West Antarctica
(see \cref{fig:so2})
and is significantly smaller and fine scale than cases above.
The floating ice sheet is approximately
$115 \textrm{km}$
long and with the water column below up to around
$1 \textrm{km}$ deep (\cref{fig:pig1}),
again with an notably acute aspect ratio.

RTopo is relatively coarse at this scale and instead we select a finer dataset
generated directly by an observational campaign using an AUV \citep[Autosub 3, built by
the UK National Oceanography Centre and deployed by the British Antarctic Survey,][]{dutrieux14}.
In this case there exists no orientated vector path of the grounding line position, nor of the coastline.
From self-consistent ice draft and bedrock positions that have been uniformly filtered,
a \brep identifying the position of the grounding/coast line is constructed using \shingle.
This high fidelity \brep is shown in \cref{fig:pig1}(a) in a local UTM plane projection,
with a relatively coarse spatial discretization
$\mathcal{T}_h$, colored by its inherited identification from $n_{\Omega'}$
which marks where sponge conditions are to be applied
in a region accurately bounded by orthodromes.
With this coarse spatial approximation, the smaller land masses are not directly represented in the boundary of the resulting geoid discretization.
The \brep has been clipped along a parallel, and then extended in local UTM space to incorporate a sponge region required for relaxing to open ocean conditions.

Full discretization of the ocean domain
is shown in \cref{fig:pig1}(b),
extruded to the
self-consistent fields of bedrock and ice draft.
Geoid spatial resolution is homogeneous at approximately
$1 \textrm{km}$.
This resolution is larger than the raw data,
so a Gaussian filter is applied throughout to all fields based on this
required local spatial resolution.
With the domain being built up from a single self-consistent dataset,
it is possible to apply this consistently to all sources used in the domain discretization,
and notably in the \brep
such that self-consistency is maintained and
the resultant discrete domain is
self-consistent.

\begin{figure}[!h]
\begin{center}
\includegraphics[width=\columnwidth]{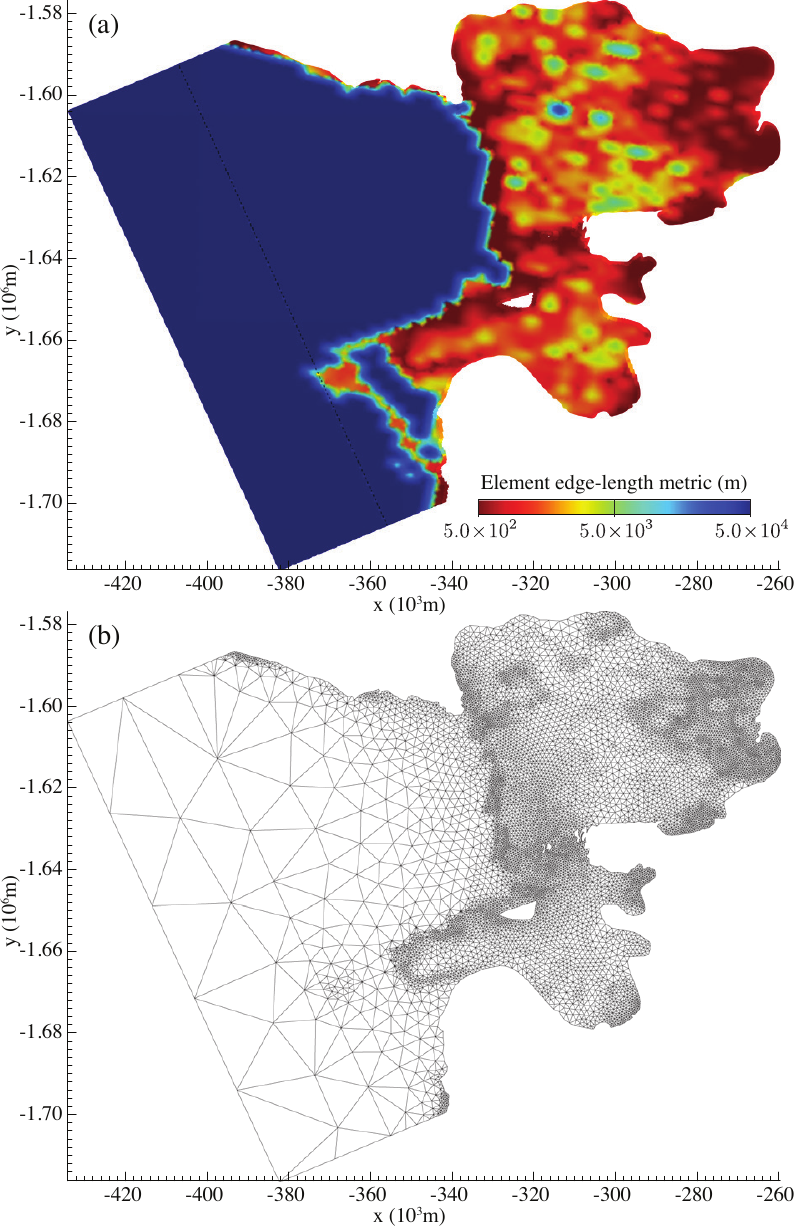}
\end{center}
\caption{
(a) Geoid element edge-length metric
$\mathcal{M}_h$ based on the local gradient of water column thickness, presented in a local UTM projection.
(b) Resulting multi-scale geoid spatial discretization
$\mathcal{T}_h$, of the Pine Island Glacier ice shelf ocean cavity.
}
\label{fig:pig2}
\end{figure}

With a relatively high spatial resolution of the source dataset, there is the possibility to optimize the geoid spatial resolution through
the metric
$\mathcal{M}_h$
\eqref{geohmetric}.
\Cref{fig:pig2} shows the scalar geoid metric field
\begin{equation*}
\mathcal{M}_h (\boldsymbol{x})
=
\mathcal{A}(
\|
\nabla (g^o(\boldsymbol{x}) - f^o(\boldsymbol{x}))
\|
),
\end{equation*}
which describes a spatial resolution based on the
local change in gradient of water column thickness.
The gradient is scaled by
$\mathcal{A}$,
a simple affine transformation,
such that the
range of scales varies from
$500 \textrm{m}$
to
$50 \textrm{km}$.
The resulting geoid discretization contains a higher spatial resolution (for both the \oned boundary and \twod surface discretizations)
at the ice front, along the deepest parts of the grounding line, and in the region close to the back where a network of geometrically complex sub-basal channels exist in the floating sheet.

\subsection{Southern Ocean}
\label{sec:southernocean}

Following discretizations in the small and geometrically complex
ice shelf ocean cavities,
this section extends this established self-consistent process out to the SO,
demonstrating the efficient prototyping, scalability and hierarchy of automation of the approach
(\cref{tenet:scales,tenet:automated,tenet:standard}).

A high fidelity \brep is produced from the \emph{amask} field of RTopo,
capturing all the land masses of Antarctica.
Here the domain contains
a large open boundary aligned along a parallel completely circumscribing the globe.
The implementation \shingle closes the domain along this free boundary and ensures this is well represented in the output discretization.
With the focus on the southern hemisphere, the stereographic projection
\eqref{stereographicprojection}
used in generating
$\mathcal{T}_h$ is made about a projection point at the North Pole instead of the South.

\begin{figure}[!h]
\vspace*{2mm}
\begin{center}
\includegraphics[width=0.9\columnwidth]{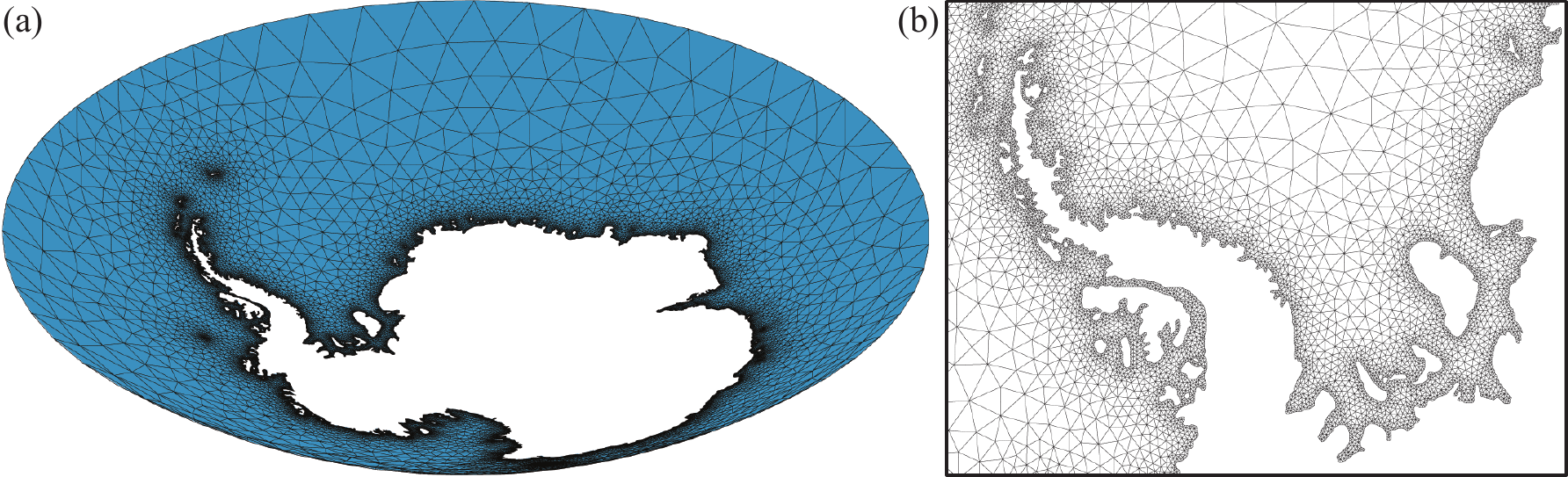}
\end{center}
\vspace{-3.8ex}
\caption{
(a) Surface geoid discretization $\mathcal{T}_h$ of the ocean domain around Antarctica,
with the open bounding parallel extended up to 50\degree S.
The computational domain includes ice shelf ocean cavities and meshing proceeds up to the grounding line,
or coastline where no floating ice is present.
(b) Zoomed in region,
under an orthographic projection,
highlighting the small boundary details of
the Larsen family and Filchner-Ronne ice shelves,
picked up from \cref{fig:fr}.
}
\label{fig:so1}
\end{figure}

\begin{figure*}[!h]
\begin{center}
\includegraphics[width=\textwidth]{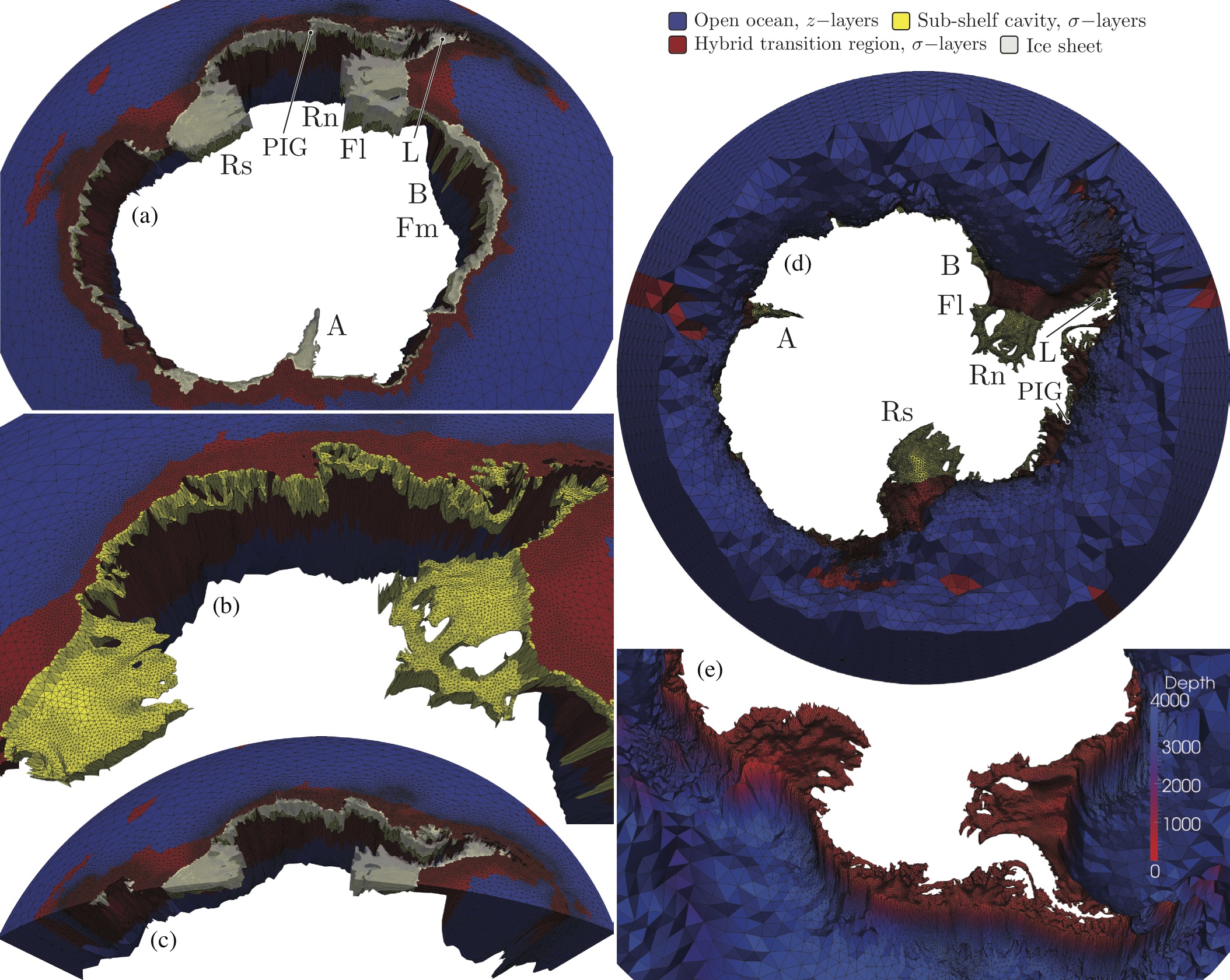}
\end{center}
\vspace{-3.8ex}
\caption{
(a)
Full discretized domain $\mathcal{T}$ of the SO up to the land and ice masses of Antarctica
is shown,
containing the largest 348 land masses of Antarctica,
together with the discretized floating ice sheet domains,
identified clockwise as
(Rs) Ross,
(PIG) Pine Island Glacier,
(Rn) Ronne,
(Fl) Filchner,
(L) Larson,
(B) Brunt,
(Fm) Fimbul,
and
(A) Amery.
These share and each conform to the ice--ocean interface surface.
(b)
Focused on the West Antarctica region,
with the discretized ice shelf domain removed,
showing the discretized full domain of the fluid ocean.
The large cavities under the Ross and Filchner-Ronne ice sheets are seen to the left and right respectively.
(c)
A vertical transect through the full discretized domain revealing the generalized hybrid coordinates in the vertical.
Region identification
$n_{\Omega'}^o$
made on the geoid marks
open ocean, continental shelf and cavity regions,
corresponding to \zlevels, hybrid transition, and \sigmalayers
colored blue, red and yellow,
respectively.
(d)
View from below highlighting the flexibility in the range of spatial scales and boundary conformity
seamlessly captured by the approach in a single multi-scale discretization,
in line with the South Pole and
(e) towards West Antarctica.
}
\label{fig:so2}
\end{figure*}

\Cref{fig:so1} shows the geoid mesh
$\mathcal{T}_h$
to this conforming bound through the process $h$ \cref{h} of \cref{fig:schematic}(b), with a proximity metric to pick up details in coast and grounding line representation.
The full discretized domains
of the ocean and floating ice sheets,
$\mathcal{T}^o$
and
$\mathcal{T}^i$
respectively
are shown in \cref{fig:so2},
with variable spatial resolution on the geoid,
and generalized hybrid vertical coordinates.

Within
$\mathcal{T}_h^o$ the region identification function
$n_{\Omega'}^o$
demarcates
the open ocean, continental shelf seas and cavities covered by a floating shelf.
Again, this is at the native resolution of the source dataset like the \brep and metric.
Generalized hybrid vertical coordinates in
$\mathcal{T}$ are developed from $n_{\Omega'}^o$, with
\zlevels in the open ocean,
\sigmalayers in the cavities
and a smooth transition between the two in the continental shelf sea region.

\label{sec:acc}

\begin{figure}[!h]
\begin{center}
\includegraphics[width=0.9\columnwidth]{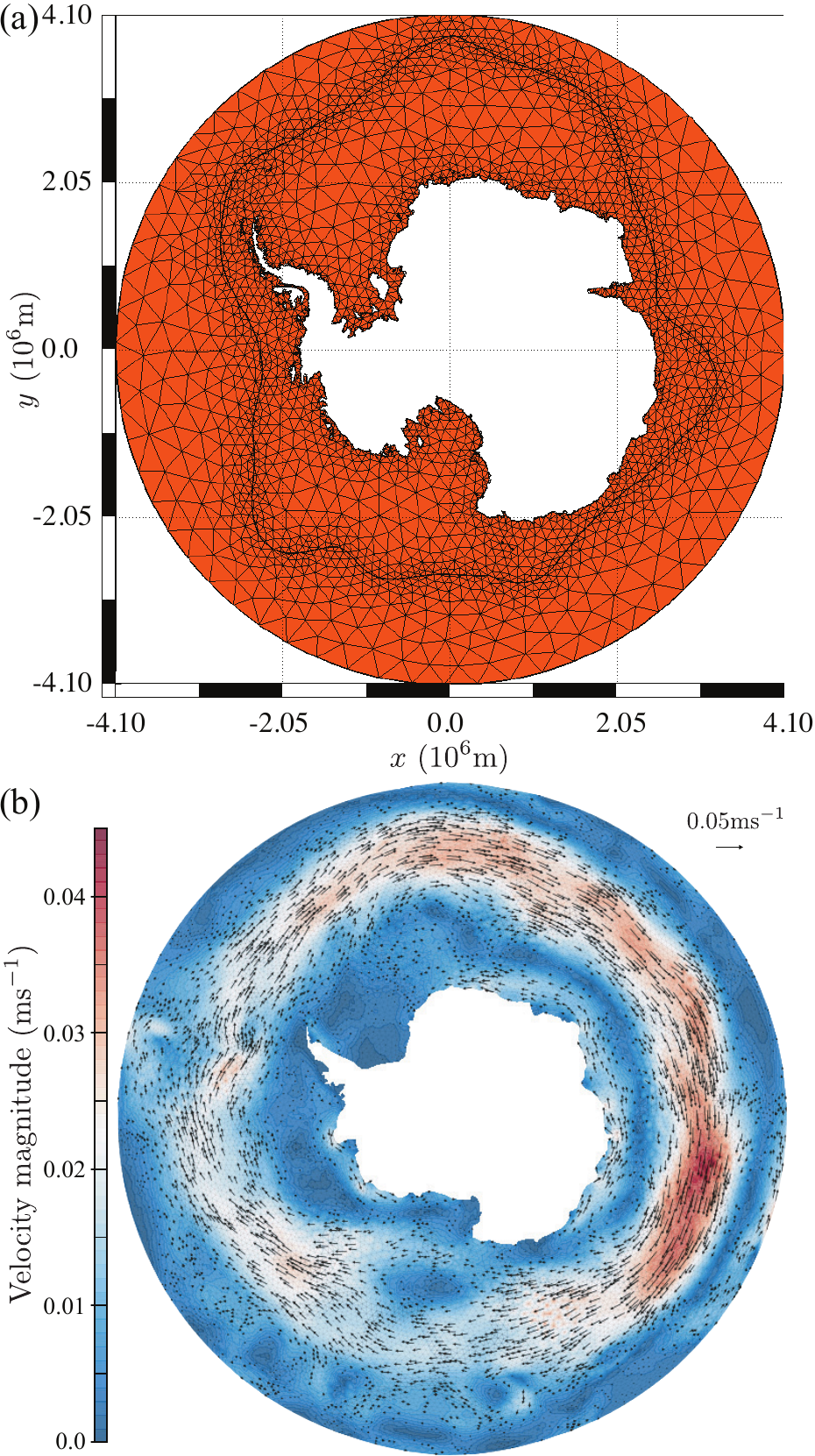}
\end{center}
\caption{
(a)
Geoid discretization
$\mathcal{T}_h$
of the ocean surrounding Antarctica,
using the same high fidelity \brep
as
\cref{fig:so1,fig:so2},
with a geoid metric
$\mathcal{M}_h$
that is a function of
the
annual mean track of the
Antarctic Circumpolar Current.
(b)
Surface velocity in the ocean surrounding Antarctica
in a simulation performed on a mesh constructed using \shingle.
Presented in a polar orthographic projection about the South Pole.
}
\label{fig:soacc}
\end{figure}

To construct the geoid discretization shown in \cref{fig:soacc}(a), a function of the annual mean track of the
Antarctic Circumpolar Current \citep[ACC, from][]{whitworth88} is trivially included in the functional providing the metric
$\mathcal{M}_h$ using the \shingle  library.
This provides a finer spatial resolution along the ACC path,
to better represent smaller-scale fluctuations in the ACC
compared to a structured, or mesh with homogeneous spatial resolution.
To ensure the coastline and grounding lines are well-represented,
the metric
is additionally a function of
proximity to these features.

\Cref{fig:soacc}(b) shows an example simulation in a SO domain constructed with this approach.
The simulation was performed
on a mesh generated from the \cite{gebco} dataset, does not include ice shelf cavities in this case,
using the
finite element model Fluidity and the
\podgpt
velocity~--~pressure element pairing
\citep{cotter09},
with the results above shown in a continuous linear space after undergoing a Galerkin projection.

\subsection{Ice sheet}
\label{sec:icesheet}

\begin{figure}[!h]
\begin{center}
\includegraphics[width=\columnwidth]{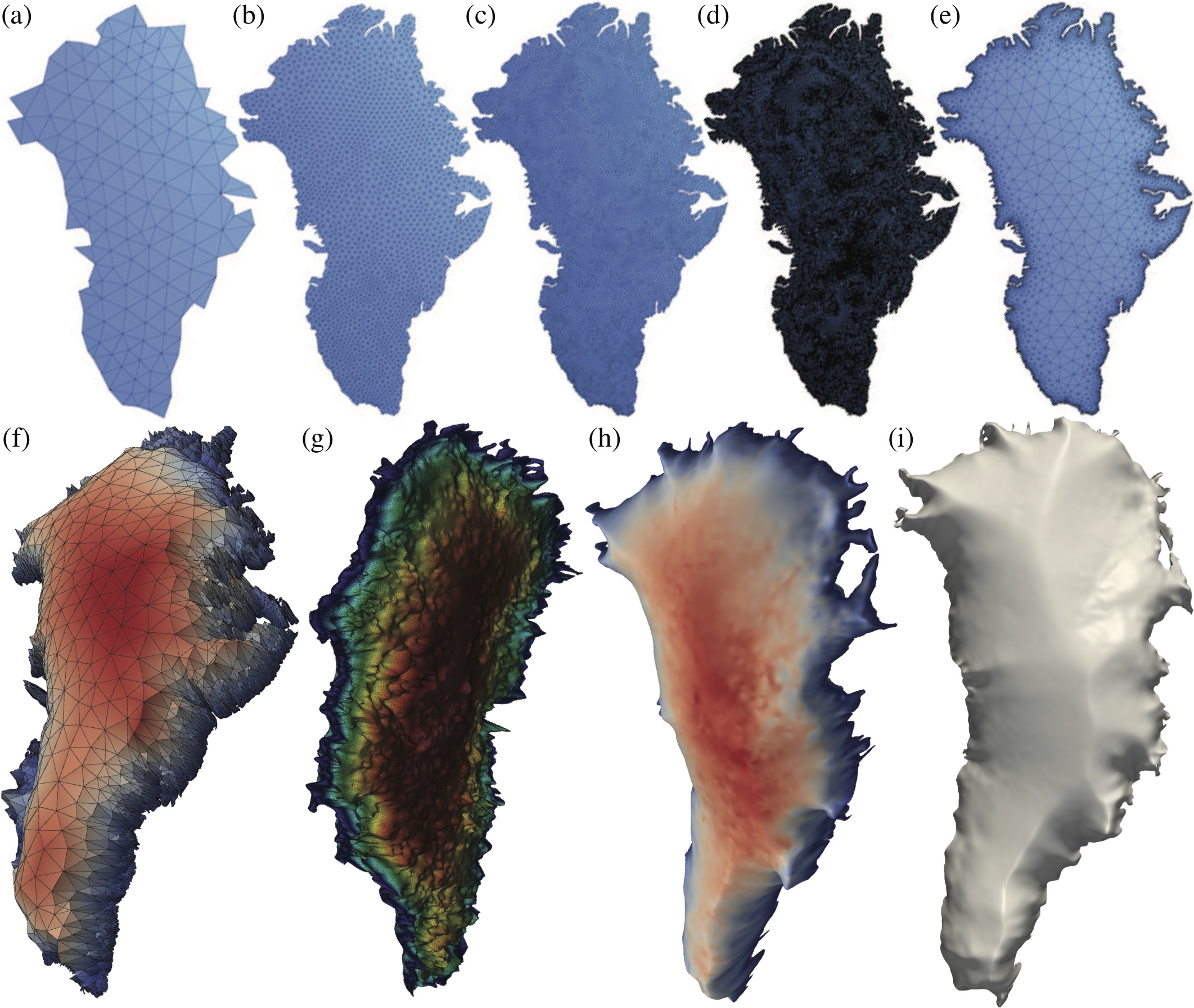}
\end{center}
\vspace{-3.8ex}
\caption{
(a)--(e) Automated geoid discretizations of the Greenland ice sheet to a terminating ice sheet thickness of 10m, at successively finer homogeneous spatial resolutions, and a multi-scale discretization ensuring the terminating front is well-represented.
Full discretizations of the Greenland ice sheet developed from the $\mathcal{T}_h$ of (a)--(e) above,
showing ice sheet thickness.
Spatial resolutions that are
(f) inhomogeneous,
(g) 5km homogeneous and 1km in
(h) and
(i),
over a meridional extent of $\sim$2,400km.
}
\label{fig:icesheet}
\end{figure}

The implementation has been applied to the Greenland and full Antarctic ice sheet.
In the case of the former, the
Greenland Standard Data Set \citep{gsd} is the source, using
the fields of
bed topography,
ice thickness
and
surface elevation
\citep[from][]{bamber01,jakobsson2012}.
High fidelity surface geoid bounds are defined
with the functional $\mathcal{F}$ in \eqref{pregeoidbrep} of the form
\begin{equation}
\mathcal{F}:=
S_d(\zeta(t))
-
S_b(\zeta(t)),
\end{equation}
where the functions
$S_d\!: \Omega' \mapsto \mathbb{R}$
and
$S_b\!: \Omega' \mapsto \mathbb{R}$
are the ice draft and bedrock bathymetry fields, respectively, from the consistently prepared source dataset.
In this case $c$ in \eqref{geoidbrep} is the terminating ice sheet thickness.

The resulting discretized \brep
at a range of spatial resolutions on the geoid plane,
constrained by
$\mathcal{M}_h$,
is shown in \cref{fig:icesheet}(a)--(e).
The full discretized domains
(\cref{fig:icesheet}(f)--(i))
were generated to bounds defined from bed topography and ice thickness fields, with an equal number of sigma layers internally, developed over multiple cores.

These spatial discretizations have been developed for simulations with a
2D Blatter-Pattyn model
written in
dolfin-adjoint with FEniCS
\citep{farrell13}
and
a 3D full Stokes model
\citep{mouradian15}.
In both cases, field data required for simulation, such as precipitation and ablation, are interpolated as finite element fields over the constructed spatial discretization,
stored and distributed efficiently in
parallel \cite{vtk} unstructured data structures.

Whilst it would be possible to use the GSHHS dataset to generate \breps of both Antarctica and Greenland that include the floating ice sheets up to the ocean interface,
it would be difficult to ensure this is consistent with other fields required, or to modify to take into account of newer datasets, or model type.
This approach has the option to easily go down to different
terminating thicknesses, depending on the simulation model and its complexity.
In the case of the single-layer 2D Blatter-Pattyn model,
for example,
this approach allowed for the efficient development of successively finer spatial approximations from
$100 \textrm{km}$
down to
$1 \textrm{km}$
resolution.
The latter containing 12,635,550 nodes and 67,253,314 triangular elements, which was used successfully to compute 2D Blatter-Pattyn model simulations on the TACC Stampede Supercomputer.

\section{Review of the nine geophysical meshing tenets}
\Cref{tenet:brep} requiring an accurate description and \emph{representation of boundaries}
$\Gamma'$
to a prescribed degree with conforming and aligned faces
is not only true of the fractal-like boundaries of the geometrically complex geophysical bounds,
but also the
smooth domain closures,
which under a piecewise-linear approximation require a minimum number of control points to be well-represented locally,
with, for example,
parallels and meridians represented well enough in stereographic space such that mesh boundaries accurately follow
orthodromes in \threed Euclidean space.

\Cref{tenet:metric} requires control over \emph{spatial resolution}, achieved through
$\mathcal{M}_h$
and
$\mathcal{M}_v$,
functions of the same self-consistent source fields as $\Gamma'$
in order to ensure the resulting spatial discretization is consistent.
From the high fidelity \brep constructed by \shingle, accurate control over the spatial discretization of the boundary is demonstrated,
and notably in the PIG case of \cref{sec:pig} where a complex geoid metric is developed based on local gradients in surface topographies.

\Cref{tenet:region} requires an \emph{accurate geometric specification of regions} which is demonstrated in the SO case of \cref{sec:southernocean}
through
$n_{\Omega'}^o$
in order to build up generalized hybrid vertical coordinates.
Additionally, boundary features are geometrically constrained, such as the sponge region aligned to parallels in the PIG discretization.

\emph{Self-consistency} in the discretized domain
$\mathcal{T}$,
covered by \cref{tenet:consistent},
is
inherited from self-consistency
present in the source data
and consistent processing of the approach.
As a result there are no issues arising from
misalignments,
no need for infilling or similar operations,
such that node positions and field values are an accurate and faithful representation of the source datasets.

The \emph{efficient drafting and prototyping} requirement of \cref{tenet:efficient} is achieved with an automated process from data to spatial discretization.
A modeler need only adjust the problem constraints, which are then faithfully adhered to by \shingle and the process \cref{fig:schematic}(b),
to give the one-one injection initially posed as the challenge in \cref{fig:challenge}.

\emph{Scalability} of \cref{tenet:scales} is demonstrated with a range of sized cases considered,
from PIG to the global oceans,
and with computationally expensive operations sent for distributed processing on HPC resources.

The process is automated, but allows individual elements of the workflow illustrated in \cref{fig:schematic}(b) to be adjusted, providing \cref{tenet:automated}, a \emph{hierarchy of automation}.
\shingle constructs a complete surface geoid domain, closing open boundaries where necessary, but permits finer scale control over constraints where required \citep[see also][]{candygis,candyshingle}.

With the injective process, and full description of constraints, the workflow is reproducible and ensures
\emph{provenance}
of the discretization development, \cref{tenet:provenance}.
Additionally, generation parameters are reproduced alongside the meshing constraints constructed by \shingle, such that generation provenance is recorded.

For the final \cref{tenet:standard},
\emph{standardization of interaction},
standard software libraries
(see \cref{sec:standard}),
geometric methods \citep[e.g.][]{gdal} and data formats (\cite{vtk} and Gmsh) are used in the approach to ensure
interoperability between both tools and scientists.
A new approach solidly handling \cref{tenet:provenance,tenet:standard} is presented in \cite{candyshingle}.

\section*{Conclusion}
\label{sec:conclusion}
This paper set out to meet four objectives to work in addressing the new and increasing challenges in taking full advantage of flexible spatial discretizations for multi-scale geophysical simulation, and in a rigorous approach.
First providing a concise, formal description of the constrain problem, which is arrived at
in \cref{sec:constraint} and specifically \cref{constraint:geophysical}.
Secondly, to outline the solution requirements for geophysical model domain discretizations, which are
detailed
in \cref{sec:tenets} and delivered in \cref{fig:tenets}.
Thirdly, to
introduce a consistent approach to the generation of \brep, geoid discretization and assembly of a full \threed discretization where necessary.
The self-consistent approach is introduced in \cref{sec:data} and detailed in
\cref{sec:brep,sec:metric,sec:id},
for the geoid \brep, spatial edge-length metric and domain identification, respectively.
For the fourth, this
enables rigorous unstructured mesh generation in general,
and specifically to accurately conform to arbitrary boundaries with only a functional definition where no orientated vector dataset exists.

This research was necessitated by the requirement to construct
boundary-conforming unstructured mesh discretizations of ice shelf ocean cavities
where it was not possible to use existing tools.
In this process the opportunity was taken to address the problem in general for geophysical models,
creating the \shingle library,
a high-level abstraction to \brep generation,
to simplify and develop an efficient method,
enabling the automated and rigorous construction of conforming boundaries to arbitrary datasets
that ensures domain consistency.

Models are advancing from simulating relatively larger-scale flows to include smaller-scale physics in a single seamless process.
These small scales bring focus to the boundaries of geophysical models, such that it is important to have accurate control over their representation and importantly, this is consistent with other simulation fields.
This demands the consideration and approach introduced,
and
is a platform for formalized, well-described and accessible routine mesh generation for unstructured mesh models.

\renewcommand*{\thesection}{\Alph{section}}
\setcounter{section}{0}%

\section{Methods to describe constraints}
\label{sec:methodstodescribe}
Relatively simple high-level constraint descriptions can be provided on the command line,
with the domain containing the Filchner-Ronne ice shelf ocean cavity shown in \cref{fig:fr}(a) straightforwardly captured with the following
\begin{align}
\texttt{\small{\  shingle }}
& \texttt{\small{-n RTopo105b\_50S.nc -f Filchner-Ronne.geo}}
\nonumber \\
& \texttt{\small{-t rtopoiceshelfcavity -lat -65.0}}
\nonumber \\
& \texttt{\small{-b -85.0:-20.0,-89.0:-75.0}}
\nonumber \\
& \texttt{\small{\phantom{-b }-67.0:-30.0,-76.0:-70.0}},
\label{command}
\end{align}
directly acting on a source file provided by \cite{rtopo}.
For cases requiring a more complex set of constraints,
the flexible, extensible approach described in \cite{candyshingle} is appropriate.
This uses
natural language,
geophysical feature based
objects in a hierarchical constraint-complete description,
that is model-independent
for sharing in general.

\section{Simulation time-varying spatial discretization}
\label{sec:timevarying}
Domain bounds and resolution metrics can be initialized from the outset to describe a discretized domain that best captures the dynamics for an entire simulation.
With a runtime adaptive algorithm, the initial domain discretization can be focused on best representing the initial conditions and coupling pathways, with the discretization then evolving in response to solution dynamics and coupling requirements.
For domains in geophysical simulations, the first time-varying extension is to allow
the resolution metrics \eqref{geohmetric} and \eqref{geovmetric} to vary in time, i.e.
${\partial\mathcal{M}_h(\boldsymbol{x}, t)}/{\partial t} \ne 0$
and
${\partial\mathcal{M}_v(\boldsymbol{x}, t)}/{\partial t} \ne 0$,
for $\boldsymbol{x} \in \Omega$
over the time interval $[0, \textrm{T})$,
whilst \cref{geobrep,geoidbound,geoidregion,geosurfbounds} remain fixed.
This redistribution of spatial resolution can be achieved with relatively efficient $r-$ and $h-$ adaptive processes.
The distinction in spatial directions
that decouples the domain discretization
needs to be preserved throughout the simulation, such that it is possible to regenerate $\mathcal{T}_h$ and then $\mathcal{T}$, evaluating the contribution of $\mathcal{M}_h$ and $\mathcal{M}_v$ through the processes \cref{h} and \cref{v} respectively.
This requires that the distinction is registered with the model code such that nodes are identified to gravitationally aligned columns, with information of the discretization $\mathcal{T}_h$ propagated to $\mathcal{T}$.
During model simulation on multiple processors, in the global ocean, ice shelf ocean cavity, ACC and SO simulations presented,
the domain mesh of is collapsed to the geoid surface mesh through an inverse prolongation operation~\citep[e.g.][]{kramer10},
and optimized with a \twod adaptive algorithm \citep[such as][]{lipnikov04}, following the evolving $\mathcal{M}_h$.
From this surface geoid mesh $\mathcal{T}_h$, the full mesh is built up through an extrusion processes to $f$ and $g$ following $\mathcal{M}_v$.

The next extension is to additionally permit the vertical bounds \eqref{geosurfbounds} to vary in time.
This is applied in the ice shelf ocean cavity simulations.
The top surface $g$, that describes the continuous ice--ocean and air-ocean boundary, adjusts in time in response to changes in pressure within the cavity,
whilst the
bottom bound $f$, describing ocean bathymetry, remains fixed.
Conservation of mass and component physical models introduce further constraints on how the domain surface is modified.
The vertical coordinate system is adjusted in time following $n_{\Omega'}$ changes, to ensure an optimal spatial discretization,
with primitives conserved following \cite{farrell09}.

The final extension is to allow the geoid plane \brep \eqref{geobrep} to vary in time.
In practice it is a computationally expensive process,
and other approaches are more efficient in adjusting
extent
on the geoid plane, such as activating regions with wetting-and-drying procedures \citep[e.g.][]{candy17}.

\subsection*{Permanence of \brepexpanded}
An arbitrary repositioning or change in node number within $\mathcal{T}_h$, with a subsequent extrusion through $f$ and $g$,
does not ensure a conservation of volume.
A simple approach which permits refinement, is to begin with a coarse $\mathcal{M}_h$ and associated $\mathcal{T}_h$, whose nodes remain in the discretization throughout a simulation,
with extrusion bounds $f$ and $g$ for any refinements determined by linear interpolation.
In this approach the representation of the surface boundary is selected at initialization and not refined during a simulation in order to conserve volume.

\section{Standardization of interaction}
\label{sec:standard}
The approach is realized in the implementation \shingle, a software library which is written in Python, a widely used high-level, interpreted programming language that has been designed to be highly extensible.
Python has a small core, with a large standard library and an easily extensible interpreter, making it easy to build up individual components in a hierarchy of automation (\cref{tenet:automated}).

It relies on established, well-regarded libraries for standard numeric operations
(including NumPy, SciPy~\citep{oliphant07} and Matplotlib)
and geospatial libraries
for robust projection and geometric operations
(including \cite{gdal}, \cite{gmtmanual}, NetCDF, shapefile, and \cite{proj4}).
Scalability built into these is inherited by the approach (\cref{tenet:scales}).
In order to quickly prototype \breps identified in \eqref{geoidbrep}, computationally demanding solutions can be cached to disk in a compact binary representation using the standard Python pickle library.

Distribution and sharing of spatial discretization constraints
for model intercomparisons, data provenance and a consistency between model setups
is addressed in \cite{candyshingle}.
Since geophysical spatial discretizations can be reproduced in an automated deterministic way using the interpreter \shingle,
it is also sufficient and constraint-complete to
depend on the self-consistent source data in a standardized format and processing record:
of pre-processing operations applied using standard common geospatial tools,
the \brep generation operation
such as \eqref{command} and metric formulae with reference to the interpretors \shingle and Gmsh.

\section*{Acknowledgments}
The \shingle library, used to develop and process the full set of \cref{constraint:geophysical} for geophysical mesh generation,
has been under continual development since 2011 and
is available to use under an open source license, with a verification test suite and example cases repeating key results shown here.
The repository is maintained at \url{http://github.com/shingleproject/Shingle},
with further information at \url{http://www.shingleproject.org}.
In addition to calculating solutions to \eqref{geoidbrep} in order to provide the consistent constraint~\eqref{geobrep},
orientated vector paths are handled directly
for use of existing, fixed paths and comparisons with other methods that develop contours with external approaches,
such as GSHHS \citep{gshhs} and GIS in \cite{candygis}.
Data used to produce the results of this paper is freely available in the cited resources and upon request to the author at
\href{mailto:a.s.candy@tudelft.nl}{\nolinkurl{a.s.candy@tudelft.nl}}.

The author wishes to
acknowledge support from
the UK Natural Environment Research Council (grant NE/G018391/1),
the Netherlands Organization for Scientific Research (NWO, grant number 858.14.061)
and thanks
Pierre Dutrieux for help with data collected by the British Antarctic Survey automated underwater vehicle \emph{Autosub} under PIG, used as the source for domain discretizations of \cref{sec:pig}.
The challenge for a generalized approach to arbitrary domain bounds was initially motivated by discussions with Matt Piggott and Paul Holland, for simulations in the complex domains of ice shelf ocean cavities.
Additionally, I thank Julie Pietrzak for helpful feedback, Patrick Farrell for testing the spatial discretizations of Greenland in the Blatter-Pattyn model development
and
Patrick Heimbach for help with computational time on the TACC Stampede Supercomputer.

\begingroup
\raggedright
\setlength{\bibsep}{0pt}
\setstretch{0.903}
\titlespacing\section{0pt}{3pt plus 1pt minus 1pt}{1pt plus 1pt minus 1pt}

\endgroup

\end{document}